\documentclass[journal,10pt]{IEEEtran}
\usepackage{amsmath,amsfonts}
\usepackage[caption=false,font=normalsize,labelfont=sf,textfont=sf]{subfig}
\usepackage{stfloats}
\usepackage{url}
\usepackage{multirow}
\usepackage{verbatim}
\usepackage{graphicx}
\usepackage{cite}
\usepackage[nolist,nohyperlinks]{acronym}
\begin{acronym}
    \acro{BPP}{binomial point process}
    \acro{CCDF}{complementary cumulative density function}
    \acro{CDF}{cumulative density function}
    \acro{PDF}{probability density function}
    \acro{BS}{base station}
    \acro{LoS}{line of sight}
    \acro{SNR}{signal to noise ratio}
    \acro{SINR}{signal to interference plus noise ratio}
    \acro{mm-wave}{millimeter wave}
    \acro{NLoS}{non line-of-sight}
    \acro{PLT}{Poisson line tessellation}
    \acro{RIS}{Reconfigurable intelligent surfaces}
    \acro{IIoT}{industrial internet of things}
    \acro{RCS}{radar cross section}
    \acro{MAC}{medium access control}
    \acro{5G}{fifth generation}
    \acro{UE}{user equipment}
    \acro{QoS}{quality of service}
    \acro{IoT}{internet of things}
    \acro{URLLC}{ultra-reliable low-latency communication}
    \acro{MAC}{medium access control}
    \acro{SISO}{single input single output}
\end{acronym}
\usepackage{xcolor}
\usepackage{float}
\usepackage{pdfpages}
\usepackage{comment}
\usepackage{siunitx}
\usepackage{graphicx,wrapfig,lipsum}
\usepackage{amsmath, amssymb, amsthm}
\usepackage{tikz}
\usepackage{verbatim}
\usepackage{url}
\usepackage[utf8]{inputenc}
\newtheorem{theorem}{Theorem}[]
\newtheorem{corollary}{Corollary}[]

\newtheorem{remark}{Remark}[]

\newtheorem{lemma}[]{Lemma}

\usepackage{multicol}

\usepackage{scalerel}
\usepackage{multicol}
\usepackage{setspace}

\usepackage{booktabs}
\usepackage{bbm}
\usepackage[normalem]{ulem}
\newcommand\blfootnote[1]{%
  \begingroup
  \renewcommand\thefootnote{}\footnote{#1}%
  \addtocounter{footnote}{-1}%
  \endgroup
}
\usepackage{color}
\usepackage{dsfont}
\usepackage{bm}
\usepackage{setspace}
\usetikzlibrary{shapes.geometric, arrows}
\allowdisplaybreaks

\usepackage{lipsum}

\usepackage{mathtools}

\makeatletter
\newcommand{\vast}{\bBigg@{3}}
\newcommand{\Vast}{\bBigg@{4}}
\makeatother

\begin{document}

\title{Spatially Correlated Blockage Aware Placement of RIS in IIoT Networks}

\author{Rashmi Kumari, Gourab Ghatak, and Abhishek K. Gupta
\thanks{R. Kumari and G. Ghatak are with the Department of Electrical Engineering, Indian Institute of Technology Delhi, India 110016. Email: \{eez228188, gghatak\}@ee.iitd.ac.in. A. K. Gupta is with the Department of Electrical Engineering, Indian Institute of Technology Kanpur, India 208016. Email:  gkrabhi@iitk.ac.in. The work is supported in part by
Qualcomm 6GUR and DST SERB grant}.}

\maketitle
\begin{abstract}
    We study the impact of deploying reconfigurable intelligent surfaces (RISs) in mitigating coverage gaps and enhancing transmission reliability in an industrial internet of things (IIoT) network. First, we consider a single blockage scenario and characterize the correlation between blocking events of the base station (BS)-user and the RIS-user links and study its impact on the probability of establishing a viable reflected link. Then, by considering multiple blockages, we derive the distribution of the signal to noise ratio (SNR) as a function of data size, blockage density, the number of RISs, and the deployment area. We analyze the impact of normalized blockage radius and identify the threshold beyond which the assumption of independent blockages deviates from the ground truth of correlated blocking. Finally, we compare the outage performance of this RIS-assisted system with that operated with network-controlled relays, and demonstrate that while the relays provide a higher reliability beyond a certain blockage threshold, increasing the number of RISs may help mitigate this effect. These insights offer valuable design guidelines for deploying RIS-aided IIoT networks in dense blockage environments.
\end{abstract}
\section{Introduction}
\label{sec:Intro}
\blfootnote{A preliminary version of this paper has been accepted for presentation at the IEEE International Symposium on Personal, Indoor and Mobile Radio Communications (PIMRC), 2025.}

Traditionally, wired communication, such as Ethernet, is common in industrial automation and monitoring setups. However, with increasing scale, it can be costly to install and maintain~\cite{cena2008hybrid}. To improve deployment flexibility and reduce costs, industries are transitioning towards wireless systems~\cite{gungor2009industrial}. Future industrial wireless networks will include various devices that demand high communication reliability and low latency~\cite{li20185g}. To meet these requirements, especially for time-sensitive control applications, \acp{URLLC} are essential in \ac{5G}, prioritizing latency, dependability, and availability~\cite{tayeb2017survey}. Additionally, massive connectivity presents challenges regarding coverage and capacity, which are addressed by protocols such as those suggested by Shah {\it et al.}~\cite{shah2019protocol}.
However, unfavorable channel conditions due to the presence of physical blockages in the factory can weaken the power of wireless signals, degrading the \ac{QoS} for such applications~\cite{cheffena2016industrial}. This effect is more prominent at high frequencies, including \ac{mm-wave} bands. To address this, Swamy et al.~\cite{swamy2015cooperative} suggested multi-hop transmission with cooperative relaying, where stronger devices help those with weaker channels. The authors in~\cite{ghatak2021stochastic} extended this by evaluating the protocol in settings with random blockages. While dedicated relays can improve network performance, their high installation costs and the complexity of optimized placement make them less practical. Instead, the study proposes leveraging existing network devices as relays, offering a scalable solution adaptable to any network size.

The deployment of \ac{RIS} have emerged as a promising approach to mitigate blockage-effects by dynamically altering the signal propagation with tunable signal reflections~\cite{mu2021intelligent}. This potentially improves coverage by providing alternate signal paths when direct links are blocked, thus mitigating blockage effects. While prior research has mainly focused on electromagnetic characterization, physical modeling, and performance assessment in simplified networks, several studies have explored deploying \acp{RIS} near transmitters or receivers to boost signal power. However, a close proximity of the \ac{RIS} to either the transmitters or the receivers may lead to a joint obstruction of the links due to correlated blocking. Consequently, simultaneous blocking of transmitter-receiver and \ac{RIS}-receiver paths can disrupt communication. 
Since blockages can affect multiple links simultaneously, leading to correlation among link blockages, it is important to study its impact on the performance of a wireless \ac{IIoT} network with \ac{RIS} to determine the optimal placement of \acp{RIS} in such a network. This paper investigates the impact of uniformly distributed \acp{RIS} in \ac{IIoT} networks under correlated blocking, focusing on addressing coverage gaps and enhancing transmission reliability in industrial applications.
\subsection{Related Work}
{\bf  On solutions to augment reliability in \ac{IIoT} Networks:} To improve the reliability of \ac{IIoT} networks, researchers have investigated strategies such as diversity techniques, relaying, and cooperative transmission~\cite{li20185g}, \cite{swamy2015cooperative}. The authors in~\cite{li20185g} have considered latency-constrained services and have proposed transmission techniques that dynamically adapt to the requirements. Whereas, the authors in~\cite{swamy2015cooperative} have demonstrated that cooperative transmission using multiple devices can lead to a high degree of reliability even in a low or moderate \ac{SNR} regime. An interesting aspect of their study is the revelation that under realistic channel conditions, merely a frequency diversity scheme is insufficient to achieve an error rate corresponding to high-reliability applications. Their proposed scheme selects a transmission rate that is characterized by the channel condition of the worst device. On the contrary, the work by Jurdi {\it et al.}~\cite{jurdi2018variable} have proposed an adaptive rate selection scheme, where, first, the channel state information of the coordinating node to the edge devices is estimated. This is followed by an adaptation of the transmission rate for each node to its instantaneous channel state. However, such a solution leads to a massive overhead due to the estimation procedure of individual channels. The authors in~\cite{brahmi2015deployment} have shown that multiple antennas reduce the \ac{SNR} required for decoding, improving the capacity and coverage of the network. The authors in \cite{ghatak2021stochastic} have demonstrated the reduction in the reliability of transmission with blockages and also evaluated the performance of multi-device cooperation schemes to alleviate the outage. However, similar to the other works on the topic, the author had largely ignored the \ac{MAC} overhead associated with such a policy. This paper proposes strategic placement of \acp{RIS} to fill coverage holes and enable high-reliability transmission for \ac{IIoT} use cases.

{\bf Impact of Blockages:} In an \ac{IIoT} scenario, the blockages can have metallic bodies which cause a detrimental impact on signal coverage, even in lower-frequency transmissions~\cite{andrews2016modeling, bai2014coverage}.  To understand the impact of blockages in such networks, stochastic geometry provides tools from random shape theory~\cite{bai2014analysis}.  Moreover, in indoor scenarios, the links may be highly correlated due to the geometric structure of the blockages and their proximity to the transmitter and the user. Most works ignore this correlation and instead assume that blocking events for different links are independent, e.g., see the recent work~\cite{okta2023blockage}. The work~\cite{gupta2022impact} showed that spatially correlated blockages can significantly impact \ac{SINR}, highlighting the need to model such effects. The impact of correlated blockages has been analyzed in 1D \ac{RIS}-assisted networks~\cite{ghatak2021deploy} and 2D \ac{mm-wave} networks~\cite{gupta2017macrodiversity,gupta2022impact}. The empirical \ac{SINR} was calculated in \cite{hriba2019correlated} using a correlation model for fixed blockages, while \cite{aditya2018tractable} studied blockage correlation for sensor localization. The authors in \cite{samuylov2016characterizing} investigated the time correlation of blocking events due to user mobility at two-time instants, and \cite{baccelli2015correlated} studied interference correlations, though simultaneous \ac{BS} blockage effects were not considered.

{\bf \ac{RIS} placement strategies:} Initial studies on \acp{RIS} have focused on physical modeling, electromagnetic characterization, and simple network performance evaluation~\cite{di2020smart}. Since the path loss of the transmitter to the receiver link via the \ac{RIS} varies as the product of the path losses of the individual (i.e., transmitter to the \ac{RIS} and the \ac{RIS} to the receiver) links, the deployment of the \ac{RIS} close to the transmitter or the receiver maximizes the received signal power~\cite{ozdogan2019intelligent}. However, such a deployment strategy may result in the transmitter-receiver and the \ac{RIS}-receiver links being blocked simultaneously, resulting in a high service outage. A comparison between dispersed \acp{RIS} and relays in \ac{SISO} systems, where users connect to the device offering the highest \ac{SNR}, has revealed that \acp{RIS} outperform relays in energy efficiency and outage probability, especially with dense deployment or high element counts \cite{ye2021spatially}. For outdoor networks, widespread \ac{RIS} placement minimizes blind spots \cite{kishk2020exploiting}. In dense environments like malls or airports, smaller, distributed \acp{RIS} provide superior coverage \cite{li2022ris}, while industrial settings often integrate \acp{RIS} on walls to enhance coverage~\cite{ren2021average}. The correlation in placements of blockages also plays a significant role in determining optimal \ac{RIS} locations in the presence of blockages, e.g.,~\cite{ghatak2021deploy} have derived correlation-aware \ac{RIS} placement strategies for outdoor \ac{mm-wave} networks. Despite their importance, the impact of indoor blockages and corresponding correlation in industrial settings remains underexplored, highlighting the lack of a comprehensive methodology for accurately analyzing \ac{RIS} performance under correlated blockage conditions.

\subsection{Contributions and Organization}
We examine the performance gains offered by the use of \acp{RIS} in an \ac{IIoT} network inside a warehouse to address coverage gaps and improve transmission reliability by taking into account the correlation among the blockages of the links involved. The key contributions of this work are summarized as follows.
\begin{enumerate}
    \item We demonstrate how uniformly deployed \acp{RIS} on the periphery of the warehouse can effectively mitigate blind spots caused by random blockages and enhance transmission reliability. For this, we consider the \ac{BS}-user link as a cascade of the direct and via-\ac{RIS} links and  characterize its \ac{LoS} probability  by considering the correlation between blocking events of these two links.  Specifically, we develop analytical methods to characterize the correlation between blocking events of the \ac{BS}-\ac{RIS} link and the \ac{RIS}-user link while determining the \ac{LoS} probability of the via-\ac{RIS} link between the \ac{BS} and the user. This analysis reveals the conditions under which a viable reflected link can be established.
    \item We extend the analysis to networks with multiple \acp{RIS} by considering the correlation between the blockages of the \ac{BS}-\ac{RIS} and \ac{RIS}-user links for each individual \ac{RIS}, while considering independent blocking across different \acp{RIS}. We derive closed-form expressions for the \ac{LoS}/\ac{NLoS} probabilities and validate them using Monte-Carlo simulations, highlighting the influence of blockage size, location, and user position on link availability. 
    \item Leveraging the derived \ac{LoS} probability, we derive the outage probability of a uniformly located user in the network. We analyze the impact of the target rate, blockage size, and blockage distance on the outage probability and identify conditions under which a given \ac{URLLC} \ac{QoS} requirement in terms of reliable delivery of a fixed payload with minimal latency can be guaranteed. Additionally, we analyze the impact of the normalized blockage size and identify the point beyond which the performance under the independent blockage assumption begins to deviate noticeably from correlation-aware models, highlighting the limitations of assuming independence in environments with larger blockages. More importantly, we highlight the importance of strategic and user-centric placement of \ac{RIS} such that it creates new paths to user and BS with mutually uncorrelated blocking events.
    \item Finally, using simulations, we show that beyond a certain number of blockages, a network-controlled relay or repeater can outperform a single \ac{RIS} deployment in terms of reducing outage probability. Our analysis shows that deploying multiple \acp{RIS} can surpass the outage performance of a network-controlled relay, offering a scalable solution for enhancing reliability under severe blockage conditions. To determine the number of \acp{RIS} needed to effectively address coverage gaps and enhance transmission reliability in \ac{IIoT} networks, we derive the corresponding \ac{SNR}-based rate coverage probability for the user. Additionally, we examine the impact of warehouse radius and the number of \acp{RIS} on outage probability under various blockage conditions. These insights provide valuable guidance for deploying \ac{RIS}-aided \ac{URLLC} wireless communication networks in \ac{IIoT} environments.
\end{enumerate}
The rest of the paper is organized as follows. In Section \ref{sec: SystemModel}, we introduce the system model for the \ac{RIS}-assisted factory environment, discuss the deployment of the \acp{RIS}, characterize the blockage process, and outline the path loss model. Section \ref{sec: NLoS} covers the derivation of the probability that the link between the \ac{RIS}-\ac{BS} and the user is in \ac{NLoS}, while Section \ref{sec: COVERAGE PROBABILITY} covers the derivation of \ac{SNR} and rate coverage probability of the user. We discuss the numerical and simulation results in Section \ref{sec: NUMERICAL RESULTS AND DISCUSSION}. Finally, Section \ref{sec: Con} concludes this paper by examining the number of \acp{RIS} needed to surpass the performance of relays.  
\section{System Model}
\label{sec: SystemModel}
\begin{figure}[t]
\centering
{\includegraphics[trim={0.7cm 0.7cm 0.7cm 0.6cm},clip,width = 0.8\columnwidth]{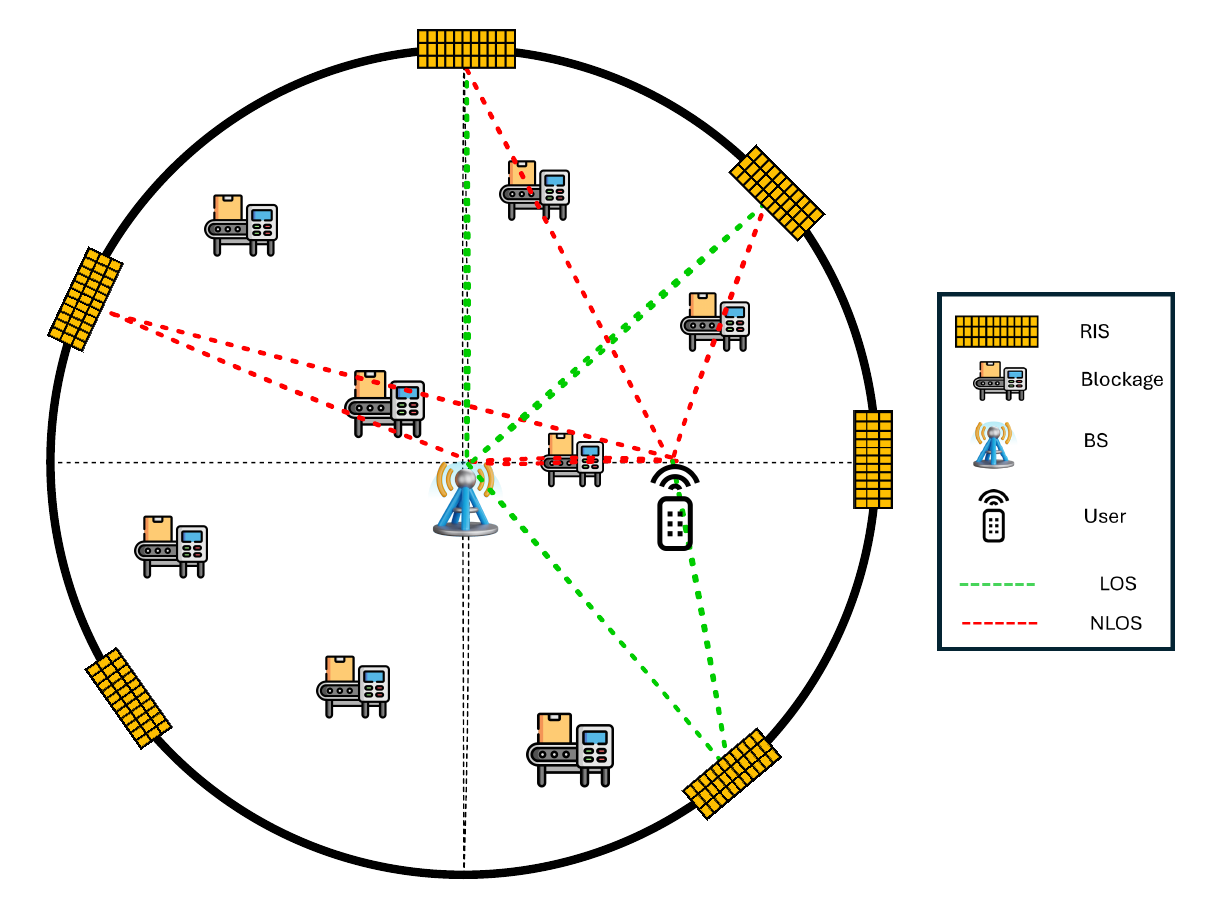}}
\caption{System Model}
\label{fig: System Model}
\end{figure}

\begin{figure}[t]
\centering
\includegraphics[trim={1.5cm 3.5cm 6.9cm 3.5cm},clip,width = 0.48\linewidth]{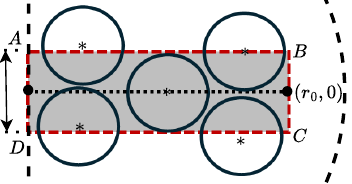}
\includegraphics[trim={2.1cm 1.4cm 0.5cm 0.1cm},clip,width=0.5\linewidth]{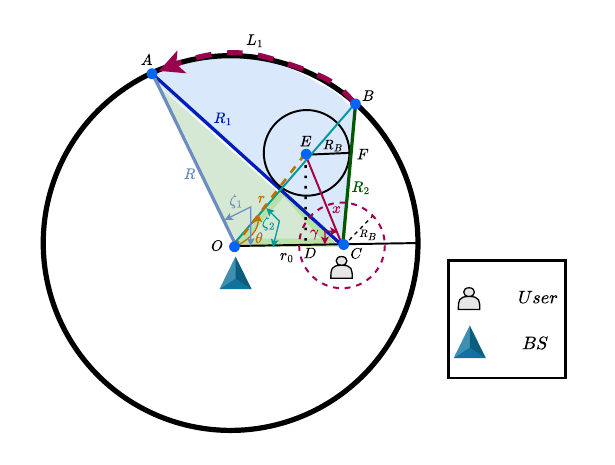}

 {\bf \small  \hspace{-1.4in}(a) \hspace{1.5in}(b)}
 \caption{An illustration showing (a) the direct link blockage and (b) the range where \ac{RIS} is blocked from the user given a blockage at $E$}\vspace{-.2in}
\label{fig: Bs User Link}
\end{figure}
\subsection{Network Geometry}
We consider an \ac{IIoT} network with a \ac{BS} placed at the center of a warehouse shaped like a circular disk with radius $R$. Let us consider a user located at a distance of $r_0$ along the positive x-axis\footnote{We assume that $r_0 \geq 2R_B$ since if the user is closer to the \ac{BS}, it is unlikely that the \ac{BS}-user link is blocked.}, as illustrated in Fig.~\ref{fig: System Model}. Due to dynamic blockages, the link between the \ac{BS} and the user can either be in \ac{LoS} or \ac{NLoS}. Assume $n_{\rm R}$ \acp{RIS}, each equipped with $M$ passively reflecting elements, uniformly deployed on the inner wall at the boundary of the room. Let the location of the $i^\text{th}$ \ac{RIS} be denoted by $\mathbf{z}_i$, where $i = 1, 2, \dots, n_R$.\\
\textbf{Blockages.} Let $n_{\rm B}$ blockages be located at $\{\mathbf{y}_j=r_{j}\angle \theta_{j}\}_{j = 1}^{n_{\rm B}}$ within the circular room such that their distances from the center is uniform and independent across blockages. Each blockage is modeled as a disk of radius $R_{\rm B}$. Their positions relative to the origin are given by ($r\cos\theta$, $r\sin\theta$), where $r \sim \mathcal{U}(R_{\rm B}, R - R_{\rm B})$ and $\theta \sim \mathcal{U}(0, 2\pi)$, where $\mathcal{U}(a,b)$ denotes the uniform distribution between $a$ and $b$. Although the blockages are modeled as circular disks, they are allowed to overlap in this work, offering a more realistic representation of industrial environments where blockages may intersect.\\
\textbf{\ac{SNR} model for the direct link.} The direct received power $P_{\rm D}$ at the user from the \ac{BS} is $P_{\rm D} = Ph_{\rm D}Kr_0^{-{\alpha_{\rm l}}}$, where $r_0$ is the distance of the user from the \ac{BS}, $h_{\rm D}$ denotes the fading power gain on the direct link. We assume Rayleigh fading in the direct link which implies that $h_{\rm D}$ is exponentially distributed.\\
\textbf{\ac{SNR} model for the indirect links.} The indirect links are the cascaded links from the \ac{BS} to the \acp{RIS} and subsequently from the \acp{RIS} to the user. The indirect received power $P_{{\rm I}_i}$ is \cite{kokkoniemi2021stochastic}
\begin{align}
  P_{{\rm I}_{i}} = P K M^2\sum_{i=1}^{n_{\rm R}} h_{{\rm I}_i} (\ell(r_0,\phi_i) R)^{-{\alpha_{\rm l}}}, 
\end{align}
where $\ell(r_0,\phi_i)$ represents the distance from the user to the $i^{\mathrm{th}}$\ac{RIS}, $R$ denotes the distance from the \acp{RIS} to the \ac{BS}, and $M$ is the total number of \ac{RIS} elements. The indirect link fading is assumed to be Nakagami-m distributed with shape parameter $m$ and spread parameter $\Omega$, which accurately captures a wide range of propagation conditions and is widely adopted in the RIS literature for analytical tractability~\cite{samuh2020performance}. 
As per Alzer’s lemma~\cite{andrews2016modeling}, the \ac{CCDF} of $h_{\rm I}$ is upper-bounded as
  $F_{h_{\rm I}}(x) \leq  \sum_{n=1}^{m}(-1)^{n+1}\binom{m}{n} e^{-nx\eta}$, 
with $\eta = m(m!)^{-\frac{1}{m}}$. Before proceeding with the analysis of the indirect link, we first recall that the blockage probability of the direct link (let us call it event $E_0$) can be derived using the void probability of a \ac{BPP}. As illustrated in Fig.~\ref{fig: Bs User Link}(a), a link obstruction occurs when the center of at least one blockage lies within the shaded region (ABCD)\footnote{Note that the actual blockage region is a capsule shape consisting of a rectangle of area $2R_B r_0$ and two half-circles of total area $\pi R_B^2$. For analytical simplicity, the circular cap area is neglected, which is valid when $R_B \ll r_0$ and $R_B \ll R$.}, which has dimensions $r_0 \times 2R_B$.
\begin{align*}
\mathbb{P}\left(E_0\right) = q_{\mathrm{B}}(r_0) = 1-\left(1-\frac{2r_0R_{\rm B}}{{\pi}{R^2}}\right)^{n_{\rm B}}.
\end{align*}
\section{Blockage Probability of the Indirect Link}
\label{sec: NLoS}
A blockage in the signal path between a user and an \ac{RIS} can disrupt \ac{BS}-\ac{RIS} and \ac{RIS}-user connections, depending on their relative location and blockage size, leading to a correlation in blocking of these links. To highlight the correlation in these links, we first consider the case of a single blockage before extending to the multiple blockages scenario.
\begin{figure}[t]
\centering
\includegraphics[trim={1.5cm 3.5cm 0.5cm 0.8cm},clip,width = 0.48\linewidth]{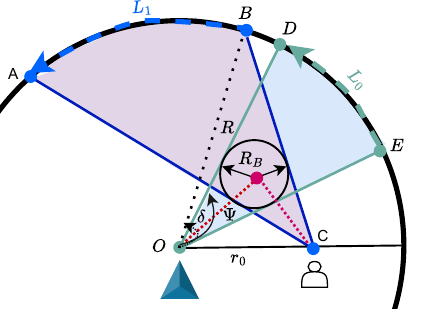}
\includegraphics[trim={1.5cm 3.5cm 0.5cm 0.8cm},clip,width = 0.48\linewidth]{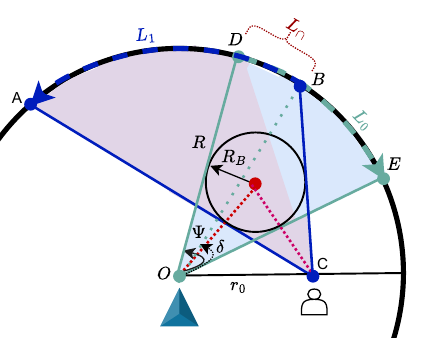}
\vspace{-.2in}

{\bf \small \hspace{-1.4in}(a) \hspace{1.5in}(b)}
\caption{An illustration showing the \ac{RIS} location range that may result in blocking of \ac{RIS}-\ac{BS} and \ac{RIS}-User link for given blockage for two cases.}\vspace{-.2in}
\label{fig: RIS-BS and RIS-User Link}
\end{figure}

\subsection{Blocking Correlation for Single Blockage}
\label{Blockage correlation}
Let us consider a blockage located at $E$ at location $\mathbf{y}=r\angle\theta$ (See Fig. \ref{fig: Bs User Link}(b)). 
\subsubsection{Blockage of RIS-user link}
Let the \ac{RIS} be located uniformly anywhere on the boundary. Let $Z_{\mathrm{R_{1}U}}$ denote the link between the \ac{RIS} and the user. Let $B_{\mathrm{R_{1}U}}$ represent the event that $Z_c$ is blocked 
 which occurs when the \ac{RIS} is inside the arc $AB$ where $AB$ is the projection of the blockage at the boundary from the user (see  Fig.~\ref{fig: Bs User Link}(b)) with its length $L_1$ given in the following Lemma.
 \begin{lemma}
 \label{lemma1}
 The length $L_1$ of $AB$ is
 \begin{align}
     L_1 = 
     \label{eq:L1}
     \begin{cases}
       2\pi R,\qquad\qquad\quad \textrm{if}\;\; (r^2 - 2rr_{0}\cos\theta)\leq R_{B}^2\! -\! r_{0}^2,\\
       {R}{\left({ \arcsin{\left(\textstyle\frac{{r_0}{\sin\gamma_2}}{R} \right)}} - { \arcsin\left({\textstyle\frac{{r_0}{\sin\gamma_1}}{R}}\right)}+ {\chi} \right)}
     ,\textrm{o.w.},
   \end{cases}
\end{align}
where $\gamma_1 = \gamma -\frac{\chi}{2}$, $\gamma_2 =\gamma +\frac{\chi}{2}$,\\ with $\gamma = \arcsin\left(\frac{r\sin\theta}{{\sqrt{(r\sin\theta)^2 + (r_0 - r\cos\theta)^2}}}\right)$,\\ and $\chi={2}\arcsin\left(\frac{R_B}{{\sqrt{(r\sin\theta)^2 + (r_0 - r\cos\theta)^2}}}\right) $.
 \end{lemma}
\begin{IEEEproof}
    From Fig.~\ref{fig: Bs User Link}(b), the arc length $L_1 = {R}{(\zeta_1 - \zeta_2)}$.
    In  $\Delta CDE$,  $\angle D = \frac{\pi}{2}$,  $CD = r_0 - r\cos\theta$, and $ED = r\sin\theta$.\\ Accordingly, $x\stackrel{\Delta}{=}CE=  {\sqrt{(r\sin\theta)^2 + (r_0 - r\cos\theta)^2}}$. \\Further, $\gamma\stackrel{\Delta}{=}\angle ECD= \arcsin\left(\frac{r\sin\theta}{x}\right)$.\\ 
    In triangle $\Delta ECF$,  $\angle ECF$ is $\frac{\chi}{2}$, where 
    Let ${\chi} \stackrel{\Delta}{=} 2\angle ECF={2}\arcsin\left(\frac{R_B}{x}\right)$. Let $\gamma_1$ and $\gamma_2$ denote $\angle ACO$ and $\angle BCO$. Applying the sine rule in triangles $\Delta AOC$ and $\Delta BOC$ respectively yields
\begin{align}
    &{\zeta_1} = {\pi - \arcsin\left(\textstyle\frac{{r_0}{\sin(\gamma_1)}}{R}\right) -\gamma_1},\nonumber\\
    \text{and }&
    {\zeta_2} = {\pi - \arcsin\left(\textstyle\frac{{r_0}{\sin(\gamma_2)}}{R}\right) -\gamma_2}.
\label{zeta_1}
\end{align}
Thus, the arc length $L_1$ simplifies to \eqref{eq:L1}.
\end{IEEEproof}
\subsubsection{Blockage of RIS-BS link}Similarly,   the event $B_{\mathrm{BR_1}}$ that the link $Z_{\mathrm{BR_{1}}}$ between the \ac{RIS} and the \ac{BS} is blocked, corresponds to \ac{RIS} being inside $DE$. Here, $DE$ is the projection of the blockage at the boundary from the \ac{BS} with its length $L_{01}$, obtained by substituting \( r_0 =0\) in $L_1$, given as
 \begin{align}
     L_{01} =R \chi_2,
     \label{L0}
 \end{align}
 where $\chi_2 = 2\arcsin\left(\frac{R_B}{r}\right)$. 
\subsubsection{Blockage of cascade link}The cascade indirect link is blocked if both the \ac{BS}-\ac{RIS} and \ac{RIS}-user links are blocked simultaneously, i.e., $B_{\mathrm{BR_{1}}} \cup B_{\mathrm{R_{1}U}}$. However, $B_{\mathrm{BR_{1}}}$ and $B_{\mathrm{R_{1}U}}$ are correlated, since the same blockage object can obstruct both links, making the computation of the joint blocking probability non-trivial. Here, blockage correlation refers to the statistical dependence between the blocking events of the \ac{BS}-\acp{RIS} and \acp{RIS}-user links, following the discussion in~\cite{gupta2017macrodiversity}. In fact, the event $B_{\mathrm{BR_{1}}} \cap B_{\mathrm{R_{1}U}}$, where both links are blocked, corresponds to the case that \ac{RIS} lies inside the intersection of $AB$ and $DE$. Its length $L_{{\cap}_{1}}$ depends on whether the arcs $AB$ and $DE$ overlap. Let $\Psi = \theta + \frac{\chi_2}{2}$ and $\delta = \zeta_2$.  If $\delta \geq \Psi$, the arcs do not intersect (see Fig.~\ref{fig: RIS-BS and RIS-User Link}(a)), otherwise the arcs intersect (See Fig.~\ref{fig: RIS-BS and RIS-User Link}(b)). Therefore, 
\begin{align}
    L_{\cap_1} = 
    \begin{cases}
        0 &\text{if } \delta \geq \Psi, \\
         R \left( \theta + \frac{\chi_2}{2}-\zeta_2 \right) &\text{otherwise},
   \end{cases}
   \label{Lcap}
\end{align}
Now, note that the cascade (via-\ac{RIS}) indirect link $Z_\mathrm{BR_{1}U}$ is \ac{NLoS} if any of the two links $Z_\mathrm{BR_{1}}$ or $Z_\mathrm{R_{1}U}$ is blocked {\em i.e.} $B_{\mathrm{R_{1}U}}\cup B_{\mathrm{BR_{1}}}$.  Assuming the \ac{RIS} location is uniformly distributed on the boundary, we get the following result. 
\begin{lemma}
  \label{lm: RIS_User}
  Given a blockage at $\mathbf{y}=r\angle\theta$, the probabilities that the link $Z_\mathrm{BR_{1}}$, the link $Z_\mathrm{R_{1}U}$ and the cascade indirect link $Z_\mathrm{BR_{1}U}$ is blocked, are respectively given as  
  \begin{align}
   &\mathcal{P}_\mathrm{R_{1}U}(r,\theta)= \frac{L_1}{{2}{\pi}{R}}, \qquad {\mathcal{P}_\mathrm{BR_{1}}}(r,\theta) =  \frac{L_{01}}{{2}{\pi}{R}},
   \label{RIS-USER}\\
   &\mathcal{P}_\mathrm{BR_1U}(r,\theta)=\mathbb{P}(B_{\mathrm{R_{1}U}}\cup B_{\mathrm{BR_{1}}})=\frac{L_1+L_{01}-L_{\cap_{1}}}{{2}{\pi}{R}}. \label{intersection}
\end{align}
\end{lemma}
\begin{IEEEproof}
    Lemma~\ref{lm: RIS_User} can be derived from Lemma~\ref{lemma1}, together with ~\eqref{L0} and ~\eqref{Lcap}.
\end{IEEEproof}
\begin{remark}
$L_{\cap_1}$ captures the correlation in the \ac{BS}-\ac{RIS} and the \ac{RIS}-user links which is generally ignored in existing literature.
\end{remark}
\begin{corollary}
  For a blockage located at $\mathbf{y} = r \angle \theta$ and $n_{R}$ \acp{RIS}, the probability that the cascade indirect link $Z_{\mathrm{BR}U}$ is blocked is given as  
  \begin{align}
   &\mathcal{P}_\mathrm{BRU}(r,\theta)=\prod_{i=1}^{n_R}\mathbb{P}(B_{\mathrm{R_{i}U}}\cup B_{\mathrm{BR_{i}}})=\prod_{i=1}^{n_R}\Bigg(\frac{L_{i}+L_{0i}-L_{{\cap}_{i}}}{{2}{\pi}{R}}\Bigg) ,\label{union}
\end{align}
where $i=1,2,\dots,n_R$. Here $L_{i}$, $L_{0i}$ and $L_{{\cap}_{i}}$ denotes the effective blockage lengths associated with the \ac{RIS}-user, \ac{BS}–\ac{RIS}, and their intersection for the $i^{\mathrm{th}}$ \ac{RIS}, respectively.
Note that for each \ac{RIS}, the \ac{BS}-\ac{RIS} and \ac{RIS}-user links are correlated, while signals from different \acp{RIS} are assumed independent. 
\end{corollary}
\subsection{User-\ac{RIS} Link Visibility With Multiple Blockages}
\label{multi}
\begin{figure}[ht!]
\centering
{\includegraphics[trim={3.5cm 3.5cm 0.5cm 0.5cm},clip,width = 0.48\linewidth]{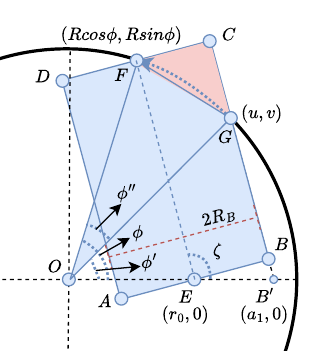}
}
\centering
\hfill
\centering
{\includegraphics[trim={17cm 17cm 0.8cm 0cm},clip,width = 0.48\linewidth]{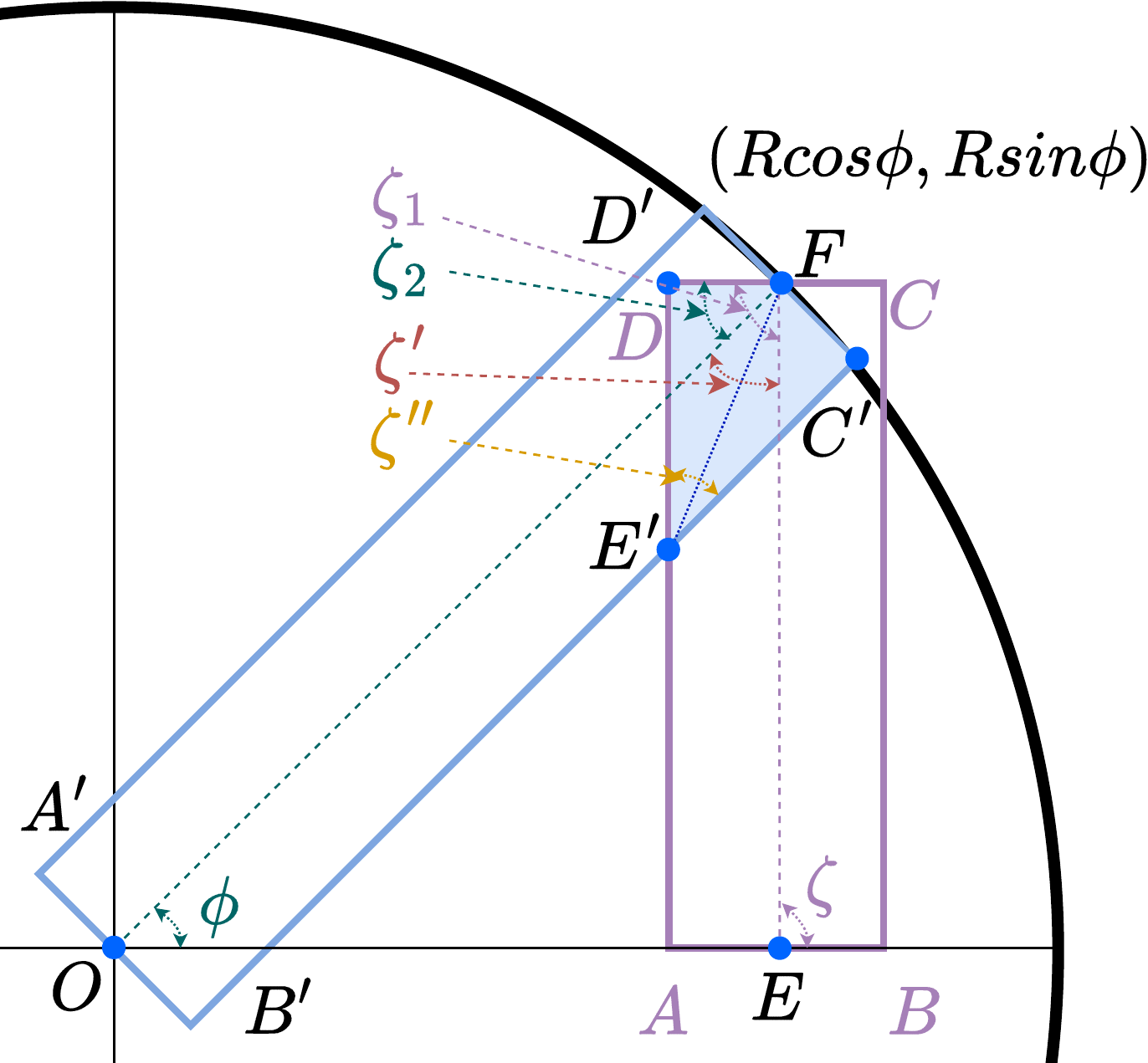}
}
\vspace{-.25in}

 {\bf \small \hspace{-1.4in}(a) \hspace{1.5in}(b)}
\caption{Illustration showing the region where blockages should not fall to ensure the visibility of (a) \ac{RIS}-User  and (b) \ac{RIS}-\ac{BS} and \ac{RIS}-User.}
\label{fig:3}
\end{figure}

To analyze the case with multiple blockages, we condition on the location of the \ac{RIS} and study the statistics across random realizations of the blockage process. As per Fig.~\ref{fig:3}(a), let an \ac{RIS} be located at point $F$ at an angle $\phi$ from the origin $O$. The user is at $E \coloneqq (r_0, 0)$. A single blockage can be approximated by a line segment of length $2R_B$ acting as its projection. Accordingly, line segments of length $R_B$ are drawn from the user towards points $A$ and $B$, and from the \ac{RIS} towards points $C$ and $D$, forming the rectangle $ABCD$ \footnote{The actual geometry would be a rounded rectangle, which is approximated by a rectangle for analytical tractability. In the numerical results section, we will show that this approximation does not affect the key insights.}. The \ac{RIS}-user link is blocked if the center of a blockage lies within this rectangle. However, since a portion of $ABCD$ extends beyond the room where blockages cannot exist. The effective region of interest is the intersection region denoted by $ABGFD$, where the blockages should not lie. Its area $A_{ABGFD}$ needs to be calculated explicitly. 
\label{subsubsec: RIS-UE}
\begin{lemma}
\label{lemma3}
Given that the \ac{RIS} is at an angle $\phi$, the probability $\mathcal{P}_{n_B}(r_0,\phi)$ that the user-\ac{RIS} link is blocked is
\begin{align}
   &q_{\mathrm{R_1U}}= \mathcal{P}_{n_B}(r_0,\phi) = 1- \left(1-\frac{A_{ABGFD}}{\pi R^2}\right)^{n_B}
   \label{f_1}\\
&\text{where } A_{ABGFD} =\nonumber\\
   &\nonumber \!{2R_B\ell_1}\! -\! \!\left( \!\!\frac{R^2\sin\phi''}{2} \!+\! \frac{R_B(\ell_1-z_1+R_B/m_1)}{2}\! -\!\frac{\phi'' R^2}{2}\!\!\right)\!. \nonumber
\end{align}
with $\ell_1 = \sqrt{(R \cos \phi - r_0)^2 + R^2 \sin^2 \phi}$ and $z_1 = \sqrt{v^2 + (u - a_1)^2}$, with $u = \dfrac{m_1^2 a_1 + \sqrt{R^2 (1 + m_1^2) - m_1^2 a_1^2}}{1 + m_1^2}$, $v = m_1 (u - a_1)$, $m_1 = \dfrac{R \sin \phi}{R \cos \phi - r_0}$, $a_1 = \frac{R_B}{m_1} \sqrt{(m_1^2 + 1)} + r_0$, and $\phi'' = \phi - \arcsin \left( \dfrac{z_1}{R} \left( \pi - \arctan(m_1) \right) \right)$.
\end{lemma}
\begin{IEEEproof}
We calculate the area of $ABGFD$ by subtracting the area of $\Delta FCG$ and the area of the segment $OFG$ from the area of the square $ABCD$. For this, we first derive the area of the triangles $\Delta FOG $, $ \Delta FCG $, and sector $FOG$, as shown in Fig.~\ref{fig:3}(a). The equation of line $ FE $ is $ y = m_1(x - r_0) $, where $ m_1 = \frac{R\sin\phi}{R\cos\phi - r_0} $. The distance between points $ F $ and $ E $, denoted by $ \ell_1 $, is $ \ell_1 = \sqrt{(R\cos\phi - r_0)^2 + R^2\sin^2\phi} $. The distance between $ E $ and $ B $ is  $ R_B = m_1\frac{|a_1 - r_0|}{\sqrt{m_1^2 +1}} $. The coordinate of $ (a_1,0) $, where line $ CB $ intersects the positive $ x $-axis at point $B'$, is $ a_1 = \frac{R_B}{m_1}\sqrt{(m_1^2+1)} + r_0 $. Similarly, the equation of line $ CB $ is $ y_1 = m_1(x - a_1) $. The coordinates of point $ G(u,v) $, where line $ CB $ intersects the disk of radius $ R $ centered at $ (0,0) $, are $ G\left(\frac{m_1^2a_1 + \sqrt{R^2(1+m_1^2)-m_1^2a_1^2}}{(1+m_1^2)},m_1(u-a_1)\right) $. Let $ z_1 $ be the distance between points $ G $ and $ B' $, given by $ z_1 = \sqrt{v^2 + (u-a_1)^2} $. Using the Sine rule in $\Delta GOB'$, $ \phi' = \arcsin\left(\frac{z_1}{R}(\pi - \zeta)\right) $. The angle of sector $ FOG $ is $ \phi'' = \phi - \phi' $, and the length $ FG $ is $ R\phi'' $. Thus, the area of sector $ FOG $ is
   $A_{FOG} = \left(\frac{\phi'' R^2}{2}\right)$.
The sides of $\Delta FOG$ are $FO = R$, $OG = R$, and the included angle $\angle FOG = \phi''$, then the area of $\Delta FOG$ is
${A_{\Delta FOG}} = \frac{R^2\sin\phi''}{2}$.
The area of the right triangle $ \Delta FCG $, with base $ FC = R_B $ and height $ CG = \ell_1 - \left(z_1 +\frac{R_B}{m_1}\right)$  is
    ${A_{\Delta FCG}} = \frac{R_B\left(\ell_1- \left(z_1 +\frac{R_B}{m_1}\right)\right)}{2}$.
The area of the polygon ${ABGFD}$ is obtained by subtracting the area of $ \Delta FCG $ the rectangle $ ABCD $, given by $ A_{ABCD} = 2R_B \ell_1 $, as shown in Fig.~\ref{fig:3}(a). 
Adding the area of sector minus $\Delta FOG$ into it, we get the desired result.
\end{IEEEproof}
Similarly, given that the \ac{RIS} is at an angle $\phi$, the probability that the \ac{BS}-\ac{RIS} link is blocked is given as
\begin{align}
   &\!\!q_{\mathrm{BR_1}}\!=\!\!\left(\!\!1\!-\!\frac{{2R_BR} \!-\! 2\left(\! \frac{R^2\sin\phi'}{2}\! +\! \frac{R_B\left(\ell_1- \left(z_2 +\frac{R_B}{m_2}\right)\right)}{2} \!-\!\frac{\phi_1' R^2}{2}\!\right) }{{\pi}{R^2}}\!\!\right)^{\!\!\! n_B}\!\!\!\!\!\!, 
\end{align}
where $z_2 = \sqrt{v_1^2 + (u_1 - a_2)^2}$, $v_1 = m_2 (u_1 - a_2)$, $u_1 = \dfrac{m_2^2 a_2 + \sqrt{R^2 (1 + m_2^2) - m_2^2 a_2^2}}{1 + m_2^2}$, $m_2 = \tan \phi$, and $a_2 =\frac{R_B}{m_2} \sqrt{m_2^2 + 1}$. Also, $\phi_1'=\phi-\phi_1$, where $\phi_1 = \arccos \left( \dfrac{R^2+a_2^2 -z_2^2}{2Ra_2} \right)$. This result can be obtained easily by substituting $r_0=0$ in Lemma \ref{lemma3}.
\subsection{Visibility of Cascade User-\ac{RIS}-\ac{BS} Link Under Multiple Blockages}
Next, we see that the shaded region $ AEB'E' $ represents the intersection between the \ac{RIS}-user link, shown in Fig. \ref{fig:3}(b). Hence, the probability that $Z_\mathrm{R_1U}$ and $ Z_{\mathrm{BR_1}}$  are blocked simultaneously is
    $\mathbb{P}{(Z_\mathrm{R_1U}\cap Z_{\mathrm{BR_1}})}  = 1- \left(1-\frac{A_{FDC'E'}}{\pi R^2}\right)^{n_B}$.
Therefore, the probability that the cascade indirect (via-\ac{RIS}) link is blocked ({\em i.e.} at least one of the \ac{BS} to \ac{RIS} link and the \ac{RIS} to user link is blocked) is given as
\begin{align}
 &\mathbb{P}{(Z_\mathrm{R_1U}\cup Z_{\mathrm{BR_1}})}  = q_{\mathrm{R_1U}}
     +q_{\mathrm{BR_1}}-1+ \left(1-\frac{A_{FDC'E'}}{\pi R^2}\right)^{n_B} \!\!\!\!\!\!\nonumber. 
\end{align}

\begin{theorem}
 \label{th: theorem2}
  The probability that the cascaded indirect (via-\ac{RIS}) link is blocked, is given as
 \begin{align}
    &q_{\mathrm{BR_1U}}=\mathbb{P}{(Z_\mathrm{R_1U}\cup Z_{\mathrm{BR_1}})}  =  q_{\mathrm{R_1U}} \\ 
     + q_{BR_1} \nonumber\\ &\! 
    -1+\left(1-\frac{1}{\pi R^2}\left(\frac{R_B^2}{\tan(\frac{\zeta"}{2})}\right)\right)^{n_B}\!\!\!\!,
\end{align}
where $\zeta''= \arctan\left(\frac{R \sin \phi}{R \cos \phi - r_0}\right)- \phi$. 
 \end{theorem}
 \begin{IEEEproof}
 In the quadrilateral $ DFC'E' $, we have $ \angle D = \angle C' = \pi/2 $ and $ \angle F + \angle E' = \pi $. Given $ DF = FC' = R_B $, two adjacent sides are equal, and one opposite angle is right, classifying $ DFC'E' $ as a kite. To validate this, drawing a bisector from $ F $ to $ E' $ shows that triangles $ \Delta FDE' $ and $ \Delta FC'E' $ are congruent by the RHS criterion. Specifically, $ \angle D = \angle C' $, both are right angles; $ DE' = C'E' $, the adjacent sides are equal, and $ FE' $ is the common hypotenuse. Thus, $ DFC'E' $ forms a kite, and $ DE' = C'E' $. The kite’s area is the sum of $ \Delta FDE' $ and $ \Delta FC'E' $. Let $ \zeta' = \zeta_1 - \zeta_2 $, where $ \zeta_1 = \arctan(m_1) $ and $ \zeta_2 = \arctan(m_2) $, with $ m_1 $ and $ m_2 $ as slopes of lines $ FE $ and $ FO $, respectively. The side length $ DE' $ is $ DE'  =  \frac{R_B}{\tan(\frac{\zeta"}{2})} $, where $ \zeta" = \zeta' $ by parallelogram properties. Hence, the area of $ \Delta FDE' $ is $ (\frac{R_B^2}{{2}\tan(\frac{\zeta"}{2})}) $. Thus, the shaded region $ FDC'E' $ area is

    $A_{FDC'E'}  = \left(\frac{R_B^2}{\tan(\frac{\zeta"}{2})}\right)$.

\end{IEEEproof}
\subsection{Visibility of Cascade User-\acp{RIS}-\ac{BS} Links Under Multiple Blockages and Multiple \acp{RIS}}
\label{subsec: multiple RIS- multiple blockages}
As discussed in Theorem~\ref{th: theorem2},  the blockage events affecting the user-\ac{RIS} and \ac{RIS}-\ac{BS} links can be fully correlated, i.e., if a blockage obstructs one of the two links, it is likely to obstruct the other as well. Further, for tractability in our analysis, the blockage events across different \acp{RIS} are considered statistically independent, as the spatial separation between \acp{RIS} leads to distinct propagation paths and non-overlapping blockage regions. This hybrid correlation model captures a realistic scenario where the joint blockage behavior within a single cascade link is preserved, while still allowing spatial diversity across multiple \acp{RIS} to mitigate the risk of complete link outage. In Section~\ref{sec: NUMERICAL RESULTS AND DISCUSSION}, we demonstrate with numerical results that this approach lends remarkable accuracy to the actual performance while maintaining a high degree of tractability. On the contrary, removing the correlation aspect in the individual \ac{RIS} links results in a significant loss of accuracy.
\begin{lemma}
\label{th: lemma4}
 Assuming that the blockage events across different \acp{RIS} are statistically independent, the probability that all cascade links via the  $n_{\rm R}$ \acp{RIS} are simultaneously blocked is given by
    $q_{\mathrm{BR}_{1:n_R}\mathrm{U}} = \prod_{i=1}^{n_R} q_{\mathrm{BR}_i\mathrm{U}}$,
where $ q_{\mathrm{BR}_i\mathrm{U}}$ denotes the probability that the cascaded link via the $ i^{\text{th}} $ \ac{RIS} is blocked and given as, $q_{\mathrm{BR_iU}}= q_{\mathrm{R_iU}}
     + q_{BR_i} \! -1+\left(1-\frac{1}{\pi R^2}\left(\frac{R_B^2}{\tan(\frac{\zeta"}{2})}\right)\right)^{n_B}$\!\!. Here $q_{\mathrm{R_iU}}= 1- \left(1-\frac{\!{2R_B\ell_i} - \left( \!\frac{R^2\sin\phi_i''}{2} \!+\! \frac{R_B(\ell_i-z_i+R_B/m_i)}{2}\! -\!\frac{\phi_i'' R^2}{2}\right)}{\pi R^2}\right)^{n_B}$  and $q_{\mathrm{BR_i}}\!=\!\!\left(\!\!1\!-\!\frac{{2R_BR} \!-\! 2\left(\! \frac{R^2\sin\phi_i'}{2}\! +\! \frac{R_B\left(\ell_i- \left(z_{2i} +\frac{R_B}{m_{2i}}\right)\right)}{2} \!-\!\frac{\phi_{1i}' R^2}{2}\!\right) }{{\pi}{R^2}}\!\!\right)^{\!\!\! n_B}$ \!\!\!. 
\end{lemma} 
\begin{IEEEproof}
Similar to Theorem~\ref{th: theorem2}, which considers the case of $n_{\rm R} = 1$, we can extend the result to multiple \acp{RIS}. The probability that the cascade indirect link via the $i^{\text{th}}$ \ac{RIS} is blocked is given by
 \begin{align*}
    &q_{\mathrm{BRU}}=\mathbb{P}{(Z_\mathrm{R_{i}U}\cup Z_{\mathrm{BR_{i}}})}  =  q_{\mathrm{R_{i}U}} \\ 
     + q_{\mathrm{BR}_{i}} \nonumber\\ &\! 
    -\mathbb{P}{(Z_\mathrm{R_{i}U}\cap Z_{\mathrm{BR_{i}}})}\!\!\!\!,
\end{align*}
where $q_{\mathrm{R}_i\mathrm{U}}$ and $q_{\mathrm{BR}_i}$ denote the probabilities that the links between the user and the $i^{\text{th}}$ \ac{RIS}, and between the \ac{BS} and the $i^{\text{th}}$ \ac{RIS}, respectively, are blocked, as discussed in subsection~\ref{multi}.
\end{IEEEproof}
 \section{\ac{SNR} Distribution} 
 \label{sec: COVERAGE PROBABILITY}
In this section, we calculate the \ac{SNR} coverage probability for a test user located at a distance $r_0$ from the \ac{BS}. We assume that the user receives the signal through reflections from only those \acp{RIS} that are in \ac{LoS} with both the user and the \ac{BS}.
The \ac{SNR} is defined as $\mathrm{SNR} = \frac{P_{\rm R}}{N_0}$, where $N_0$ is the noise power and $P_{\rm R}$ denotes the received power at the user given by $P_{\rm R} = \mathbb{I}_{\rm D}P_{\rm D} + \sum_{i=1}^{n_R} {\mathbb{I}_{{\rm I}_i} }P_{{\rm I}_i}$, where $\mathbb{I}_{\rm D}$ and $\mathbb{I}_{{\rm L}_i}$ are the indicator functions denoting the \ac{LoS} availability of the direct and the $i^\mathrm{th}$ indirect links, respectively. The direct power is $P_{\rm D} = P K h_{\rm D} r_0^{-{\alpha_{\rm l}}}$ while the reflected power is $P_{{\rm I}_{i}} = P K M^2\sum_{i=1}^{n_{\rm R}} h_{{\rm I}_i} (\ell(r_0,\phi_i) R)^{-{\alpha_{\rm l}}}$. Based on the visibility state (\ac{LoS} vs \ac{NLoS}) of the direct link, there will be only two cases: one with the direct link in \ac{LoS} and the other with the direct link in \ac{NLoS}. The communication link is considered successfully covered when $\mathrm{SNR}\geq \gamma_{\mathrm{th}}$, where $\gamma_{\mathrm{th}}$ is the minimum \ac{SNR} threshold required for the target application.  It is given by $\gamma_{\mathrm{th}} = 2^{\frac{b}{\beta T W}}$, where $b$ is the data size in bits, $\beta$ is the time partitioning factor (assumed $\beta=1$),  $T$ is the transmission time, and $W$ is the bandwidth~\cite{ghatak2021stochastic}. The \ac{SNR} coverage probability for a user located at a distance $r_0$ from the \ac{BS} is
\begin{align}
   \mathcal{P}_C(r_0,\phi_i) = \mathcal{P}_{c_1}(\gamma_{\mathrm{th}}|r_0)(1-q_{\mathrm{B}}(r_0))+\mathcal{P}_{c_2}(\gamma_{\mathrm{th}}|r_0)q_{\mathrm{B}}(r_0), 
\end{align}
where $\mathcal{P}_{{ C}_1}(\gamma_{\mathrm{th}}|r_0)$ denotes the conditional \ac{SNR} coverage probability for the case when the direct link is in \ac{LoS}, and $\mathcal{P}_{{ C}_2}(\gamma_{\mathrm{th}}|r_0)$ represents the conditional \ac{SNR} coverage probability for the case when the direct link is in \ac{NLoS}. Here, $q_{\mathrm{B}}(r_0)$ denotes the probability that the direct link is in \ac{NLoS}.
The constaint $\mathrm{SNR}\geq \gamma_{\mathrm{th}}$ is equivalent to  
\begin{align}
 \mathbb{I}_{\rm D}P_{\rm D} \geq \gamma_{\mathrm{th}}N_0 -PKM^2\sum_{i=1}^{n_{\rm R}}h_{{\rm I}_i}(\ell(r_0,\phi_i)R)^{-{\alpha_{\rm l}}}\mathbb{I}_{{\rm L}_i},
 \label{eq1}
\end{align}
where $\ell(r_0,\phi_i)= \sqrt{(R \cos \phi_i - r_0)^2 + R^2 \sin^2 \phi_i}$ denotes the distance between the user and the $i^{\mathrm{th}}$ \ac{RIS}.

\subsection{\ac{SNR} in the presence of \ac{LoS} Direct Link}
First, let us condition on the event $\mathbb{I}_{\rm D} = 1$ for a given $r_0$; then~\eqref{eq1} can be modified as
\begin{align}
 P_{\rm D} \geq \gamma_{\mathrm{th}}N_0 -PKM^2\sum_{i=1}^{n_R}h_{{\rm I}_i}(\ell(r_0,\phi_i)R)^{-{\alpha_{\rm l}}}\mathbb{I}_{{\rm L}_i}.
 \label{eq2}
\end{align}
Let the right-hand side of~\eqref{eq2} be denoted by $\mathcal{T} = \gamma_{\mathrm{th}}N_0 -PKM^2\sum_{i=1}^{n_R}h_{{\rm I}_i}(\ell(r_0,\phi_i)R)^{-{\alpha_{\rm l}}}\mathbb{I}_{{\rm L}_i}$. In order to characterize the \ac{SNR} coverage probability, first we derive the Laplace transform (LT) of $\mathcal{T}$.
\begin{lemma}
  For a given $r_0$, the LT of $\mathcal{T}$ is
  \begin{align}
    &\mathcal{B}_{\mathcal{T}}(s) = e^{-N_0\gamma_{\mathrm{th}}s}\times\; \nonumber \\&\!\left(\int_{0}^{2\pi}\left(1 - \frac{ sPKM^2(\ell(r_0,\phi)R)^{-{\alpha_{\rm l}}}\Omega}{m}\right)^{-m}\frac{\mathcal{P}_{L}(r_0,\phi)}{2\pi}\mathrm{d}\phi\right)^{n_R}\!\!\!\!\!\!,
    \label{T1}
  \end{align}
  where $\mathcal{P}_{L}(r_0,\phi)$ represents the probability that the indirect link between the \ac{BS}, an \ac{RIS}, and the user is in \ac{LoS}, and is given by $\mathcal{P}_{L}(r_0,\phi) =(1-q_{\mathrm{BR_1U}}(r_0,\phi))$.
\end{lemma}
\begin{IEEEproof}
The LT $\mathcal{B}_{\mathcal{T}}(s)$ of $\mathcal{T}$ can be expressed as the product of the corresponding transforms of the noise power $N_0$ (which is a constant) and the reflected signal power, as follows  
\begin{align}
&\!\!\mathcal{B}_{\mathcal{T}}(s) \!=\!\mathbb{E}\!\!\left[e^{- s\gamma_{\mathrm{th}}N_0 }\exp\!\left(\!sPKM^2\sum_{i=1}^{n_R}h_{{\rm I}_i}(\ell(r_0,\phi_i)R)^{-{\alpha_{\rm l}}}\mathbb{I}_{{\rm L}_i}\!\!\right)\!\!\right]\!\!\nonumber\\
\!\!&=\!e^{-N_0s\gamma_{\mathrm{th}}}\;\!\mathbb{E}\left[\exp\left(sPKM^2\sum_{i=1}^{n_R}h_{{\rm I}_i}(\ell(r_0,\phi_i)R)^{-{\alpha_{\rm l}}}\mathbb{I}_{{\rm L}_i}\!\right)\!\right]\!\!\nonumber\\
&=e^{-N_0s\gamma_{\mathrm{th}}}\;\mathbb{E}\left[\prod_{i=1}^{n_R}e^{\left(sPKM^2h_{{\rm I}_i}(\ell(r_0,\phi_i)R)^{-{\alpha_{\rm l}}}\mathbb{I}_{{\rm L}_i}\right)}\right]\nonumber\\
&=\mathcal{B}_{N_0}(s\gamma_{\mathrm{th}})\mathcal{B}_{P_{\rm I}}(s),
 \label{eq3}
\end{align}
where $\mathcal{B}_{N_0}(s\gamma_{\mathrm{th}})= e^{-N_0s\gamma_{\mathrm{th}}}$ and $\mathcal{B}_{P_{\rm I}}(s) = \mathbb{E}\left[\prod_{i=1}^{n_R}e^{\left(sPKM^2h_{{\rm I}_i}(\ell(r_0,\phi_i)R)^{-{\alpha_{\rm l}}}\mathbb{I}_{{\rm L}_i}\right)}\right]$. Further, $\mathcal{B}_{P_{\rm I}}(s)$ can be simplified as 
\begin{align}
&\mathcal{B}_{P_{\rm I}}(s) =\mathbb{E}\left[\prod_{i=1}^{n_R}e^{\left(sPKM^2h_{{\rm I}_i}(\ell(r_0,\phi_i)R)^{-{\alpha_{\rm l}}}\mathbb{I}_{{\rm L}_i}\right)}\right],\nonumber\\
&=\mathbb{E}_{\phi_i}\left[\prod_{i=1}^{n_R}\mathbb{E}_{h_{{\rm I}_i}}\left[e^{\left(sPKM^2h_{{\rm I}_i}(\ell(r_0,\phi_i)R)^{-{\alpha_{\rm l}}}\mathbb{I}_{{\rm L}_i}\right)}\right]\right].
 \label{eq4}
\end{align}
Since $h_{R}\sim\mathsf{Gamma}(m,\Omega/m)$, then the moment-generating function (MGF) of $h_{R}$ is given by
  \begin{align}
  \mathbb{E}\left[e^{u_i h_{{\rm I}_i}}\right] = \left(1 - \frac{ u_i\Omega}{m}\right)^{-m}\!, \quad \text{for }\; u_i < \frac{m}{\Omega},
  \label{eq5}
\end{align} 
where $u_i = { sPKM^2(\ell(r_0,\phi_i)R)^{-{\alpha_{\rm l}}}\mathbb{I}_{{\rm L}_i}}$.
Hence,~\eqref{eq4} can be modified as
\begin{align}
&\!\!\mathcal{B}_{P_{\rm I}}(s) =\prod_{i=1}^{n_R}\mathbb{E}_{\phi_i}\!\left[\!\left(\!1\! -\! \frac{ sPKM^2(\ell(r_0,\phi_i)R)^{-{\alpha_{\rm l}}}\mathbb{I}_{{\rm L}_i}\Omega}{m}\right)^{-m}\right].\!\nonumber\\&\!\!\text{Hence, } \mathcal{B}_{\mathcal{T}}(s)=\nonumber\\
&e^{-N_0s\gamma_{\mathrm{th}}} 
\prod_{i=1}^{n_R}\mathbb{E}_{\phi_i}\!\!\left[\!\left(\!1\! - \frac{ sPKM^2(\ell(r_0,\phi_i)R)^{-{\alpha_{\rm l}}}\mathbb{I}_{{\rm L}_i}\Omega}{m}\!\right)^{-m}\right]\!\nonumber\\
&=e^{-N_0s\gamma_{\mathrm{th}}}\;\left(\int_{0}^{2\pi}\left(1 - \frac{ u\Omega}{m}\right)^{-m}\frac{\mathcal{P}_{L}(r_0,\phi)}{2\pi}\mathrm{d}\phi\right)^{n_R}\!\!\!.
 \label{eq6}
\end{align}
\end{IEEEproof}
We now compute the SNR coverage probability of the user in the following two Theorems.
\begin{theorem}
\label{th: 2}
 The \ac{SNR} coverage probability of the user located at a distance $r_0$, when both the direct and reflected links are in \ac{LoS}, is given by
 \begin{align}
   \mathcal{P}_{c_1}(\gamma_{\mathrm{th}}|r_0)&=\mathcal{B}_{\mathcal{T}^+}\left(\frac{1}{KPr_0^{-{\alpha_{\rm l}}}}\right),
 \end{align}
 where, $\mathcal{T}^+ \triangleq \max\{0,\mathcal{T}\}$. 
\end{theorem}
\begin{IEEEproof}
The proof follows a similar approach to~\cite{sun2023stochastic}. From definition,
  \begin{align}
    \mathcal{P}_{c_1}(\gamma_{\mathrm{th}}|r_0)& =\mathbb{P}\left( KPh_{\rm D}r_0^{-{\alpha_{\rm l}}} \geq \mathcal{T}|r_0\right),\nonumber\\
   &=\mathbb{P}\left( h_{\rm D} \geq \frac{\mathcal{T}}{KPr_0^{-{\alpha_{\rm l}}}}|r_0\right).\nonumber
   \end{align}
Note that
   \begin{align}
   &\mathbb{P}\left[h_{\rm D}\ge\frac{\mathcal{T}}{KPr_o^{-\alpha_1}}|\mathcal{T},r_o\right]
   \overset{(a)}{=}\begin{cases} 
     \exp\left(-\frac{\mathcal{T}}{KPr_0^{-{\alpha_{\rm l}}}}\right) & \text{if}\;\;{\mathcal{T}>0}, \\ 
     1 & \text{if}\;\;{\mathcal{T}\leq0},
     \end{cases}\nonumber
     \end{align}
where  $(a)$ follows from the CCDF of the exponential distribution. Let $f_\mathcal{T}(\cdot)$ denote the PDF of $\mathcal{T}$. Hence, 
\begin{align}
     \mathcal{P}_{c_1}(\gamma_{\mathrm{th}}|r_0)&\overset{}{=}\int_{0}^{\infty}e^{-\frac{v}{KPr_0^{-{\alpha_{\rm l}}}}}f_{\mathcal{T}}(x)\mathrm{d}x + \int_{-\infty}^{0}f_{\mathcal{T}}(x)\mathrm{d}x,\nonumber\\
     &\overset{(b)}{=} \mathcal{B}_{\mathcal{T}^+}\left(\frac{1}{KPr_0^{-{\alpha_{\rm l}}}}\right),
     \label{eq7}
\end{align}
where $(b)$ follows from the definition of the LT of $\mathcal{B}_{\mathcal{T}^+}(s)=
 \int_{-\infty}^{\infty} e^{-s \max(t,0)} f_{\mathcal{T}}(t) \mathrm{d}t
=
\int_{0}^{\infty}e^{-sv}f_{\mathcal{T}}(x)\mathrm{d}x + \int_{-\infty}^{0}f_{\mathcal{T}}(x)\mathrm{d}x$, evaluated at the argument $s=\frac{1}{KPr_0^{-{\alpha_{\rm l}}}}$. 
\end{IEEEproof}

Note that $\mathcal{B}_{\mathcal{T}^+}(s)$ can be computed as 
\begin{align}
    \mathcal{B}_{\mathcal{T}^+}(s)\!=\!\frac{1}{2\pi i}\int_{-\infty}^{\infty}\!\!\mathcal{B}_{\mathcal{T}}(s-iw)\!-\mathcal{B}_{\mathcal{T}}(-iw)\frac{\mathrm{d}w}{w} \!+ \frac{1+\mathcal{B}_{\mathcal{T}}(s)}{2}.
    \label{eq8}
\end{align}
The  case when  the direct link is in \ac{LoS} and the reflected links are in \ac{NLoS} can be seen as special case of Theorem~\ref{th: 2} with $\mathbb{I}_{{\rm I}_i} = 0, \forall i$. Here, \eqref{eq1} can be written as
\begin{align*}
 P_{\rm D} \geq \gamma_{\mathrm{th}}N_0.
 \label{eq14}
\end{align*}
Hence, we get the following result.

\begin{corollary}
 \label{lemma: 6}   
The \ac{SNR} coverage probability for a test user located at a distance $r_0$ from the \ac{BS}, when the direct link is in \ac{LoS} and the reflected links are in \ac{NLoS} is given by
 \begin{align}
     \mathcal{P}_{c_{1}}'(\gamma_{\mathrm{th}}|r_0)\!=e^{-\frac{\gamma_{\mathrm{th}} N_0}{KPr_0^{-{\alpha_{\rm l}}}}}.     
\end{align}
\end{corollary}
\begin{IEEEproof}
This easily follows as shown below.
    \begin{align}
    &\mathcal{P}_{c_{1}}'(\gamma_{\mathrm{th}}|r_0)= \mathbb{P}\left( KPh_{\rm D}r_0^{-{\alpha_{\rm l}}} \geq \gamma_{\mathrm{th}} N_0 |r_0\right), \nonumber\\
   &=\mathbb{P}\left( h_{\rm D} \geq \frac{ \gamma_{\mathrm{th}} N_0}{KPr_0^{-{\alpha_{\rm l}}}}|r_0\right)
   \overset{}{=} 
     \exp\left(-\frac{\gamma_{\mathrm{th}} N_0}{KPr_0^{-{\alpha_{\rm l}}}}\right).
     \nonumber
    \\[-.5in]\nonumber
\end{align}
\end{IEEEproof}

\subsection{\ac{SNR} in the absence of \ac{LoS} Direct Link}
Next, let us consider the case where $\mathbb{I}_{\rm D} = 0$  for a given $r_0$. Hence,~\eqref{eq1} can be written as
\begin{align}
\underbrace{PKM^2\sum_{i=1}^{n_R}h_{{\rm I}_i}(\ell(r_0,\phi_i)R)^{-{\alpha_{\rm l}}}\mathbb{I}_{{\rm L}_i}}_{X_{{\rm L}_i}} \geq TN_0.
\label{eq12}
\end{align}
\begin{theorem}
\label{th: 3}
   The \ac{SNR} coverage probability for the test user when the direct link is in \ac{NLoS}, is
   \begin{align}
   \mathcal{P}_{c_{2}}(\gamma_{\mathrm{th}}|r_0)&=\frac{1}{2} +\frac{1}{\pi}\int_{0}^\infty\frac{\operatorname{Im}\left( e^{-it\gamma_{\mathrm{th}}N_0} \, \varphi_L(t) \right)}{t} \, \mathrm{d}t,
   \label{eq13}
\end{align}
\!here $\varphi_{L}(t)\! =\!\!\left(\!\!\int_{0}^{2\pi}\!\!\left(\!1 \!- \!\frac{ jtPKM^2(\ell(r_0,\phi)R)^{-{\alpha_{\rm l}}}\Omega}{m}\!\right)^{-m}\!\frac{\mathcal{P}_{L}(r_0,\phi)}{2\pi}\mathrm{d}\phi\!\right)^{n_R}$\!\!\!\!\!\!.
\end{theorem}
\begin{IEEEproof}
   The characteristic function of $X_{{\rm L}_i}$  is derived as follows:
\begin{align}
    \varphi_{{\rm L}}(t)&= \mathbb{E}[e^{itX_{{\rm L}_i}}]=\mathbb{E}\Bigg[e^{itPKM^2\sum_{i=1}^{n_R}h_{{\rm I}_i}(\ell(r_0,\phi_i)R)^{-{\alpha_{\rm l}}}\mathbb{I}_{{\rm L}_i}}\Bigg]\nonumber\\
    &=\mathbb{E}\left[\prod_{i=1}^{n_R}e^{\left(itPKM^2h_{{\rm I}_i}(\ell(r_0,\phi_i)R)^{-{\alpha_{\rm l}}}\mathbb{I}_{{\rm L}_i}\right)}\right]\nonumber\\
    &=\prod_{i=1}^{n_R}\mathbb{E}\left[e^{\left(itPKM^2h_{{\rm I}_i}(\ell(r_0,\phi_i)R)^{-{\alpha_{\rm l}}}\mathbb{I}_{{\rm L}_i}\right)}\right].
    \label{eq18}
\end{align}
For a given $r_0$, ~\eqref{eq18} be modified as 
\begin{align}
   & \varphi_{L}(t)=\prod_{i=1}^{n_R}\mathbb{E}\Bigg[\left(1 - \frac{ jtPKM^2(\ell(r_0,\phi)R)^{-{\alpha_{\rm l}}}\mathbb{I}_{{\rm L}_i}\Omega}{m}\right)^{-m}\Bigg],\nonumber\\
    &\!\!=\!\left(\!\int_{0}^{2\pi}\!\!\left(\!1 \!-\!\textstyle \frac{ jtPKM^2(\ell(r_0,\phi)R)^{-{\alpha_{\rm l}}}\Omega}{m}\!\right)^{-m}\!\frac{\mathcal{P}_{L}(r_0,\phi)}{2\pi}\mathrm{d}\phi\!\right)^{n_R}\!\!\!\!\!\!.
 \label{eq17}
\end{align}
We compute $\mathbb{P}(X_{{\rm L}_i}\geq \gamma_{\mathrm{th}} N_0)=1-\mathbb{P}(X_{{\rm L}_i}\leq \gamma_{\mathrm{th}} N_0)$ using the Gil-Pelaez inversion theorem~\cite{gil1951note}, as derived in Theorem~\ref{th: 3}.
\end{IEEEproof}
Therefore, based on Theorem~\ref{th: 2} and Theorem~\ref{th: 3}, the \ac{SNR} coverage probability for a test user located at a distance $r_0$ from the origin, within a circular warehouse of radius $R$, with a \ac{BS} positioned at the origin and \acp{RIS} deployed along the boundary wall at uniformly distributed azimuthal angles, is given by 
\begin{align}
    \mathcal{P}_{c}(r_0)=\mathcal{P}_{c_{1}}(\gamma_{\mathrm{th}}|r_0)\left(1-q_{\mathrm{B}}(r_0)\right)+\mathcal{P}_{c_{2}}(\gamma_{\mathrm{th}}|r_0)q_{\mathrm{B}}(r_0).\nonumber
    \end{align}
    Hence the performance of the typical user randomly located in room is given as
    \begin{align}
        &\mathcal{P}_{c}= E[\mathcal{P}_{c}(r_0)]\nonumber \\
        &=\frac{1}{R-2R_B}\int_{R_B}^{R-R_B}\Bigg( \mathcal{B}_{\mathcal{T}^+}\left(\frac{1}{KPr_0^{-{\alpha_{\rm l}}}}\right)\left(1-q_{\mathrm{B}}(r_0)\right)+\nonumber\\&\Bigg(\frac{1}{2} +\frac{1}{\pi}\int_{0}^{\infty}\frac{\operatorname{Im}\left( e^{-it\gamma_{\mathrm{th}}N_o} \varphi_L(t) \right)}{t}\mathrm{d}t\Bigg)q_{\mathrm{B}}(r_0)\Bigg)\mathrm{d}r_0.
    \label{SNR}
    \end{align}
The outage probability $\mathcal{P}_{out}(r_0, \phi_i)$ is derived from the coverage probability $\mathcal{P}_{C}(r_0, \phi_i)$ as
\begin{align}
\mathcal{P}_{out}(r_0, \phi_i)\!=\!1\!-\!\mathcal{P}_{C}(r_0, \phi_i).
\label{SNR8}
\end{align}
As data size $b$ increases,  $\gamma_{\mathrm{th}}$ rises, leading to a higher outage. Blockages further degrade \ac{SNR} and introduce variability, especially for nearby users. While \ac{RIS} can mitigate this by establishing a \ac{LoS} link, its effectiveness depends on its placement relative to the user and blockages.
\section{Numerical Results And Discussion} 
 \label{sec: NUMERICAL RESULTS AND DISCUSSION}
This section presents numerical findings to illustrate the impact of different settings on the system's performance along with validating our analysis. The base station transmits with a power of $46$ dBm\footnote{Typical for large scale industrial applications that deply macro cells rather than small cells for enterprise networks.} and an antenna gain of $17$ dBi. The warehouse radius is assumed to be either $100$ m or $200$ m. The carrier frequency is assumed to be $3.5$ GHz with bandwidth of $1$ MHz. The path loss constant is $K = \left(\frac{c}{4\pi f_c}\right)^2$, where $c$ is the speed of light, and the path loss exponent is ${\alpha_{\rm l}} = 2$. The noise power is given by $N_0 = -174$ dBm/Hz , and the  blockage radius $R_B$ is assumed to be either $0.2$~m or $0.5$~m, as considered in~\cite{ghatak2021stochastic}.
\subsection{Blockage Probability}
\label{subsec: Blockage probability}
\begin{figure}
\centering
{\includegraphics[width=0.9\linewidth]{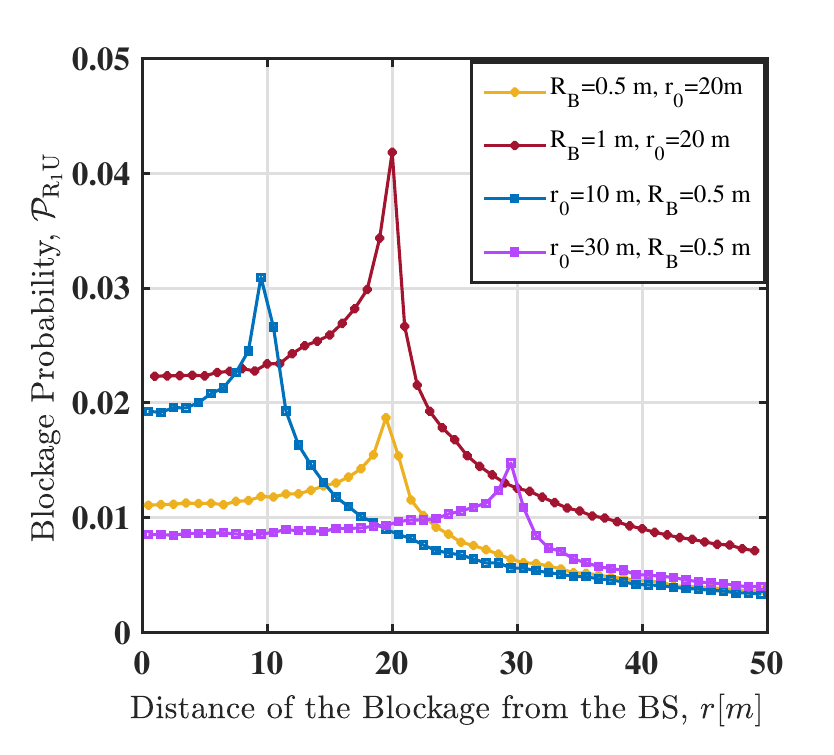}  
    \label{fig: r_user}}
\caption{$\mathcal{P}_\mathrm{R_1U}(r_0,\theta)$ of blockage of \ac{RIS}-user link, versus distance of the blockage from the \ac{BS}, $r$. The first two plots show results for different blockage radii $R_{\rm B}$ with a fixed user location $r_0$, while the remaining plots show results for different $r_0$ with a fixed $R_{\rm B}$. Solid lines represent analytical expressions, while the markers represent simulation results.}
\label{fig:4}
\end{figure}
\textbf{Blockage probability between user and \ac{RIS} in presence of single blockage}: 
Fig.~\ref{fig:4} illustrates the probability $\mathcal{P}_{\rm R_1U}$ that the user-\ac{RIS} link is blocked as a function of blockage distance $r$, with $R=50$ m for the case with single blockage as considered in Section~\ref{Blockage correlation}. The first two plots show how varying blockage radius $R_{\rm B}$ affects the blockage probability for a fixed user location $r_0=20$ m. When the blockage is near the \ac{BS} and far from the user, the blockage probability is low. As the blockage moves outward to a radial distance comparable to the user’s distance ($r \approx r_0$), the probability increases, peaking when the blockage has the highest chance of overlapping the direct path between the \ac{BS} and the user. Beyond this point, the probability decreases as the blockage moves further away. A larger $R_{\rm B}$ raises the overall blockage probability across all distances.The last two plots show the effect of varying the user location $r_0$ for a fixed blockage radius $R_{\rm B} = 0.5$ m.  Here too, the blockage probability peaks when the blockage is near the user, with the peak shifting right as $r_0$ increases, indicating that the highest blockage probability aligns with the user’s position relative to the \ac{BS}.\\
\begin{figure}
    \centering
    \subfloat[]
    {\includegraphics[width=0.85\linewidth]{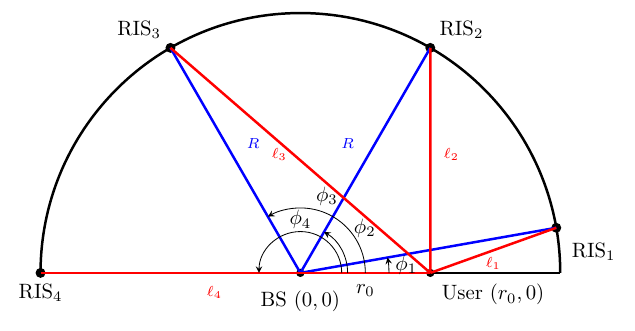}    
    \label{fig:RISorientation}}
    \hfil
    \centering
    \subfloat[]
    {\includegraphics[width=0.85\linewidth]{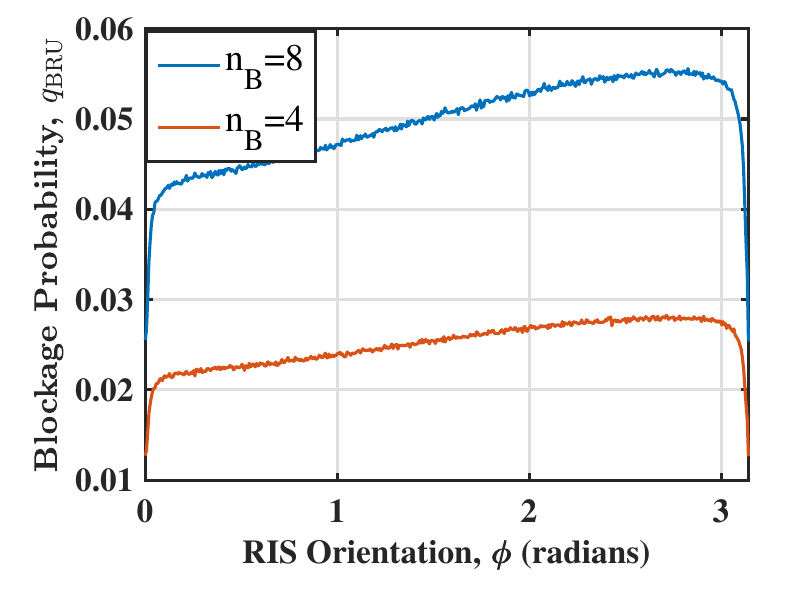}
    \label{fig: r_b_u}}
    \caption{ (a) Variation of \ac{RIS}-user path length ($\ell_i$) with the orientation ($\phi_i$) of  the \acp{RIS}. (b) Probability $q_\mathrm{BR_1U}$ that \ac{BS}-\ac{RIS}-user link is blocked versus the orientation of the \ac{RIS} $\phi$, for $R = 100$ m, $R_B = 0.2$ m, and $r_0 = 20$ m with multiple blockages.}
    \label{fig: Blockage probability BS-RIS}
\end{figure}
\textbf{Blockage probability of the cascade \ac{BS}-\ac{RIS}-user link}: Fig.~\ref{fig:RISorientation} shows the variation of the \ac{RIS}-user path length $\ell_i$ with the \ac{RIS} orientation angle $\phi_i$. When $\phi_i$ is small, the \ac{RIS} is closer to the user’s direction, resulting in a shorter $\ell_i$. As $\phi_i$ increases, the \ac{RIS} moves farther around the semicircular boundary, which lengthens the \ac{RIS}-user distance. Specifically, at $\phi = 0$, the path length is minimum, while at $\phi = \pi$, the \ac{RIS} is located on the opposite side, and $\ell_i$ attains its maximum value. Thus, $\ell_i$ increases monotonically with $\phi_i$. Fig.~\ref{fig: r_b_u} shows the variation of correlated blockage probability for the \ac{BS}–\ac{RIS}–user link with \ac{RIS} orientation angle $\phi$, for $n_{\rm B} = 4$ and $n_{\rm B} = 8$ for the system as described in Section~\ref{Blockage correlation}. When $\phi$ is small, the \ac{RIS} is near the user, so the \ac{RIS}–user path is short and and the overlap between the \ac{BS}–\ac{RIS} and \ac{RIS}–user blockage regions is large, increasing the joint blocking probability. As $\phi$ increases, the paths between the \ac{BS}, \ac{RIS}, and user get longer and the overlap between the two blockage regions gradually reduces, leading to a peak and then a drop in the correlated blocking probability. Near $\phi = \pi$, the \ac{RIS} is on the opposite side of the user, so the two paths are longest but spatially separate, minimizing their overlap and hence the joint blockage probability. With more blockages in the environment, such as when $n_{\rm B} = 8$, the overall probability of blockage is higher across all orientations. This indicates that both the \ac{RIS} orientation and the number of blockages significantly impact the reliability of the \ac{BS}–\ac{RIS}–user link. Indeed for fixed user locations, our framework thus provides a strategy to optimally deploy the \acp{RIS} in a blockage-aware manner.
\subsection{Effect of Blockage Radius on \ac{RIS} Correlation Models}
\label{subsec: blockages correlation}
\begin{figure}
    \centering
    \includegraphics[width=0.95\linewidth]{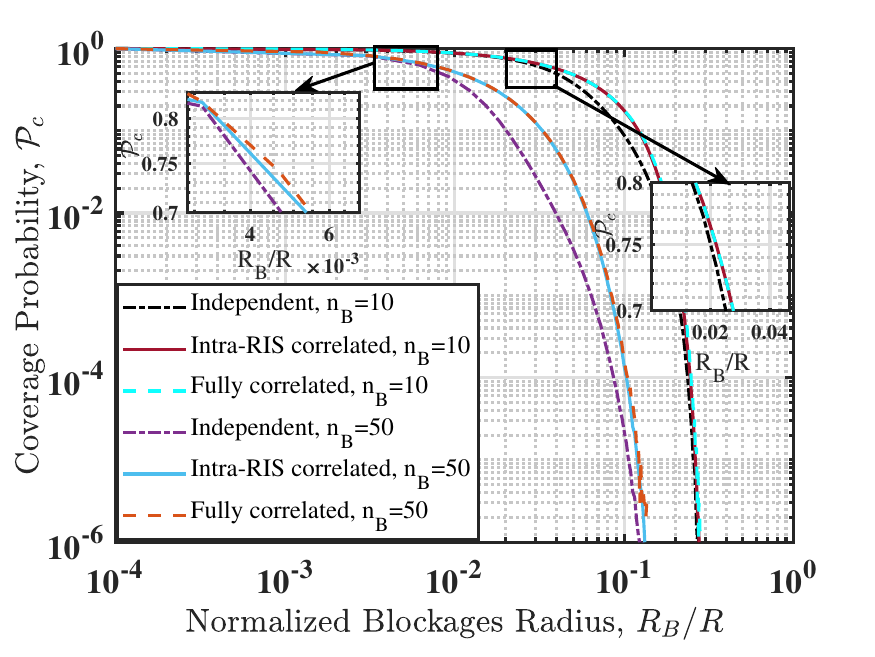}
    \caption{$\mathcal{P}_{\text{c}}$ versus normalized blockage radius $R_B / R$ for three \ac{RIS} correlation scenarios: independent, intra-\ac{RIS} correlated, and fully correlated links, with $R = 50\,\text{m}$.}
    \label{fig: correlation}
\end{figure}
Fig.~\ref{fig: correlation} illustrates the variation of coverage probability with respect to the normalized blockage size $R_B/R$, where the system radius is fixed at $R = 50$~m. The results are generated using a Monte Carlo simulation of the exact blockage model. Three blockage correlation models are compared: independent, intra-\ac{RIS} correlated, and fully correlated. In the intra-\ac{RIS} correlated model, blockages that affect the links through the same \ac{RIS} are assumed to be correlated, while links through different \ac{RIS} remain independent. In the fully correlated model, all links in the system are considered jointly affected by the same blockage events. Two blockage densities are considered: a low density with $n_{\rm B} = 10$ and a high density with $n_{\rm B} = 50$. For small blockage sizes, all three models yield high coverage probabilities, and their curves nearly overlap, especially in the zoomed-in regions. In the low-density case, intra-\ac{RIS} and fully correlated models behave similarly, indicating that correlation has minimal impact under sparse and low-blockage conditions. As the blockage size increases to a moderate level (e.g., $R_B/R = 0.02$ to $0.04$), the independent model reports lower coverage probability relative to the correlated models. This is because independent blocking events can block multiple links separately, increasing the chance that at least one link is blocked. In contrast, correlated models may block entire groups of links together, preserving some paths and improving coverage. When the blockage size becomes large (e.g., $R_B/R > 0.05$), we observe that the intra-\ac{RIS} model offers a tractable yet accurate middle ground by capturing correlation among links sharing the same \ac{RIS} without assuming full dependence. Notably, the deviation between independent and correlated models becomes apparent around $R_B/R \approx 0.025$, especially at high blockage density ($n_{\rm B} = 50$). This highlights that neglecting blockage correlation can lead to inaccurate performance estimates. Thus, Fig.~\ref{fig: correlation} demonstrates that both blockage density and correlation significantly affect performance, and accurate modeling is essential for evaluating \ac{RIS}-aided networks under realistic blockage conditions.
\subsection{Effect of the Data Size}
\label{subsec: Data size}
\begin{figure*}
    \centering
    \subfloat[]
    {\includegraphics[width=0.33\linewidth]{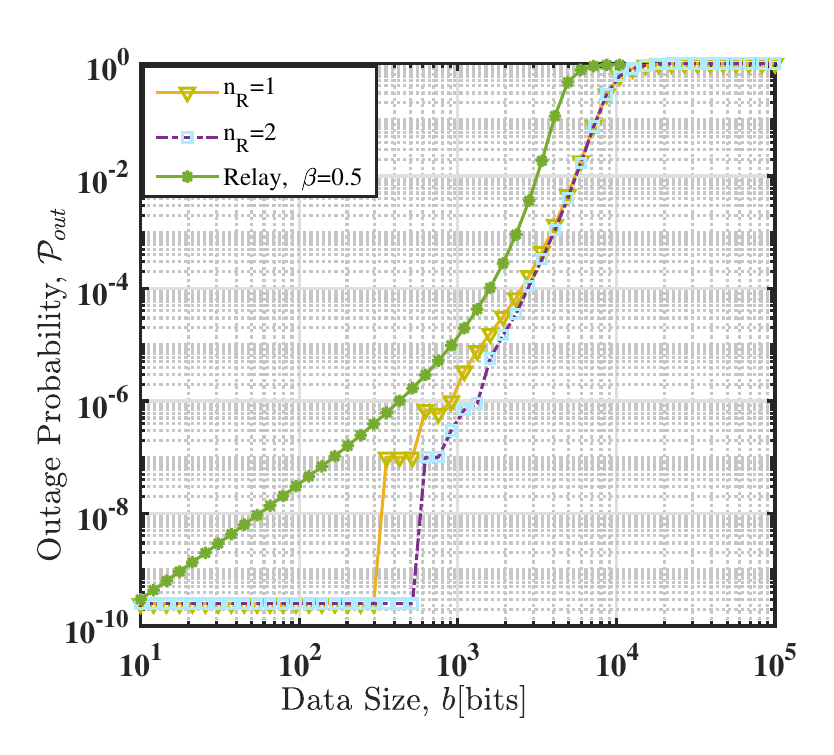}    
    \label{fig:n_B=0}}
    \hfil
    \subfloat[]
    {\includegraphics[width=0.315\linewidth]{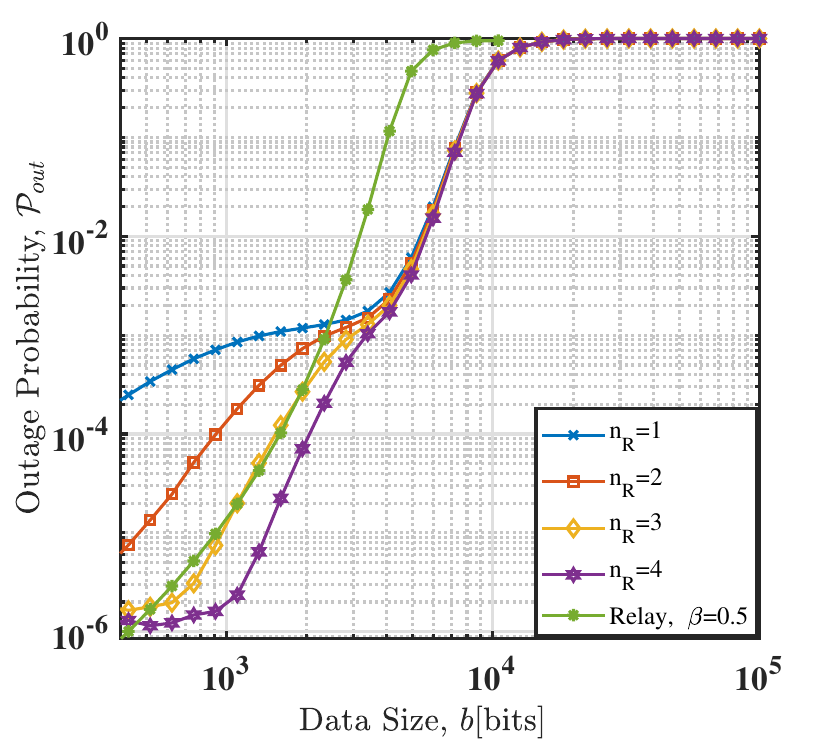}
    \label{fig:beta=0.5}}
    \hfil
    \subfloat[]
    {\includegraphics[width=0.33\linewidth]{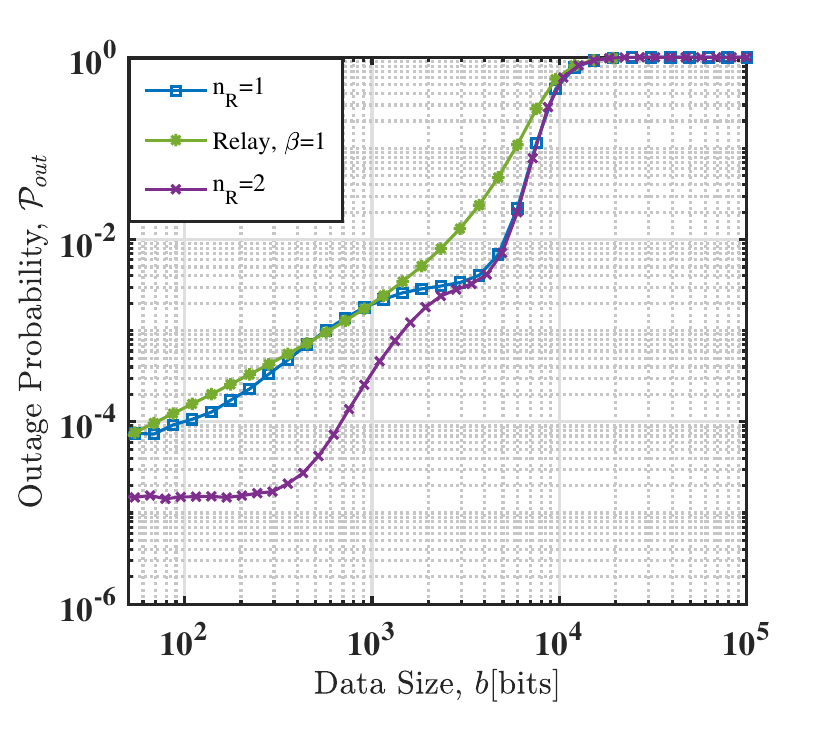}
    \label{fig:beta=1}}
    \caption{Comparison plot of the outage probability, $P_{out}$, versus data size, $b$, for different numbers of \acp{RIS} and relays; (a) with the time partitioning factor $\beta = 0.5$, considering  $n_{\rm B} = 0$; (b) with the time partitioning factor $\beta = 0.5$, considering  $n_{\rm B} = 1$ and $R_B=0.2$ m; (c) with the time partitioning factor $\beta = 1$, considering  $n_{\rm B} = 1$ and $R_B=0.5$ m, respectively.}
    \label{fig: data size}
\end{figure*}
In Fig.~\ref{fig:n_B=0}, we compare the outage probability versus data size for three different schemes: 1) an \ac{RIS} with $n_{\rm B}=0$ (hence, the \ac{BS}-\ac{RIS}-user link is always \ac{LoS}); 2) two \acp{RIS} with $n_{\rm B}=0$ and a network with; 3) relays based on the full coordination strategy as described in~\cite{ghatak2021stochastic} with no blockage and the time partitioning factor\footnote{$\beta$ controls the time split between the broadcast and relaying phases for the relay link. For the RIS, which passively reflects the signal without forwarding, we have $\beta = 1$.} $\beta=0.5$,indicating equal time allocation between the relay and broadcast phases. In all scenarios, the user-\ac{RIS}-\ac{BS} link is guaranteed to be in the \ac{LoS}, as there are no blockages. For low data sizes, the required \ac{SNR} threshold $\gamma_{\mathrm{th}}$ is relatively low, allowing the \ac{RIS} to frequently achieve the necessary \ac{SNR}, which results in a low outage probability. The outage probability with \ac{RIS} remains almost constant up to moderate data sizes (around $b=500\;\text{bits}$), while that of the relay increases gradually even at small data sizes due to its dependence on coordinated transmissions. As data size grows, the required \ac{SNR} threshold rises, resulting in a steeper increase in outage probability for both \ac{RIS} and the relay. This confirms that a single \ac{RIS} outperforms relays when no blockages are present.\\  
Fig.~\ref{fig:beta=0.5} and \ref{fig:beta=1} show the outage probability versus data sizes when both relays and \ac{RIS} configurations operate under the same parameters with $R_B = 0.2$m, $R_B = 0.5$m, and time partitioning factor $\beta =0.5$, $\beta =1$ (broadcast only), respectively. Fig.~\ref{fig:beta=0.5} shows that for smaller data sizes, the relay achieves better coverage and a lower outage probability compared to \ac{RIS} ($n_R = 1$ and $n_R = 2$, $n_R = 3$ is comparable, and $n_R = 4$ outperform the relay), whereas for larger data sizes, the \ac{RIS} provides better coverage performance than the relay. At low data sizes, single \ac{RIS} gives a higher outage probability than the relays due to its passive nature.  However, adding more \acp{RIS} increases the received signal power at the user, reducing the outage probability and narrowing the gap. In Fig.~\ref{fig:beta=0.5}, the outage probability is consistently lower than the relay for all data sizes with 4 \acp{RIS}. In Fig.~\ref{fig:beta=1}, for small $b$, the $n_R = 2$ RIS configuration achieves lower outage than the broadcast phases, while $n_R = 1$ performs similarly. For large $b$, the $n_R = 2$ \acp{RIS} continues to outperform the broadcast phases. This indicates that deploying 2 or 4 \acp{RIS} can achieve comparable or even better coverage than the relay under these specific conditions, especially when the target data size is large.

\subsection{Effect of the Number of blockages}
\label{subsec: Number of blockages}

\begin{figure}[ht!]
    \centering
    \includegraphics[width=0.8\linewidth]{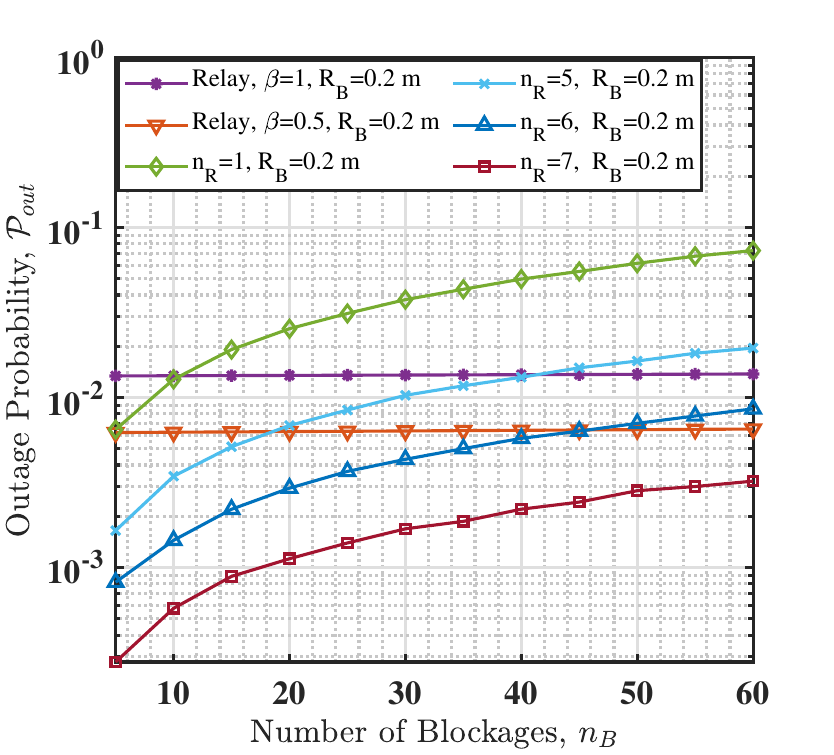}
    \caption{Comparison plot of the outage probability, $P_{out}$, versus the number of blockages, $n_{\rm B}$, for different numbers of \acp{RIS} $(n_R = 1, 5, 6, 7)$, $R=200 m$ and $R_B=0.2 m$ with relays at $\beta = 0.5$ and $\beta = 1$.}
    \label{fig: n_B}
\end{figure}

In Fig.~\ref{fig: n_B}, we plot the outage probability as a function of the number of blockages $(n_B)$ for both the \ac{RIS} configurations (with $\beta = 1$), and for the relay under two different time partitioning factors $\beta = 0.5$ and $\beta = 1$. For small $n_{\rm B}$, the user-\ac{RIS}-\ac{BS} link remains in \ac{LoS}, ensuring better coverage and low outage probability. As $n_{\rm B}$ increases, the likelihood of blockages blocking the \ac{BS}-\ac{RIS}-user link rises, leading to higher outage probability. However, at high $n_{\rm B}$, the outage probability saturates since additional blockages have minimal effect once the link is already blocked. With relays, the outage probability increases more gradually, whereas for \ac{RIS}, it rises sharply due to the lack of signal amplification.  As a result, the \ac{RIS} curve is steeper than the relay scenario. A single \ac{RIS} plot shows a higher outage probability compared to the relays. However, as the number of \ac{RIS} increases, the plot starts to flatten or decrease in the outage probability. In Fig.~\ref{fig: n_B}, the outage probability is consistently lower than that of the relay with $\beta = 1$ or surpasses the relay with 6 \acp{RIS}, whereas the relay with $\beta = 0.5$ is observed for 7 \acp{RIS}.

\subsection{Effect of the Radius of the Deployment Region}
\label{subsec: Radius of the Area}
\begin{figure*}
    \centering
    \subfloat[]
    {\includegraphics[width=0.33\linewidth]{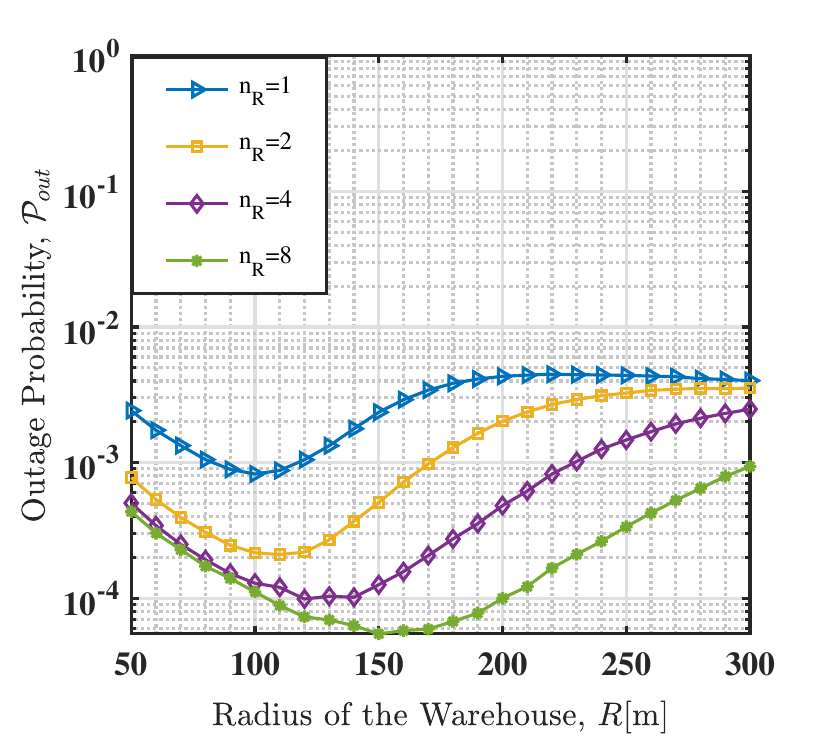}
    \label{fig: R(2)}}
    \hfil
    \subfloat[]
    {\includegraphics[width=0.32\linewidth]{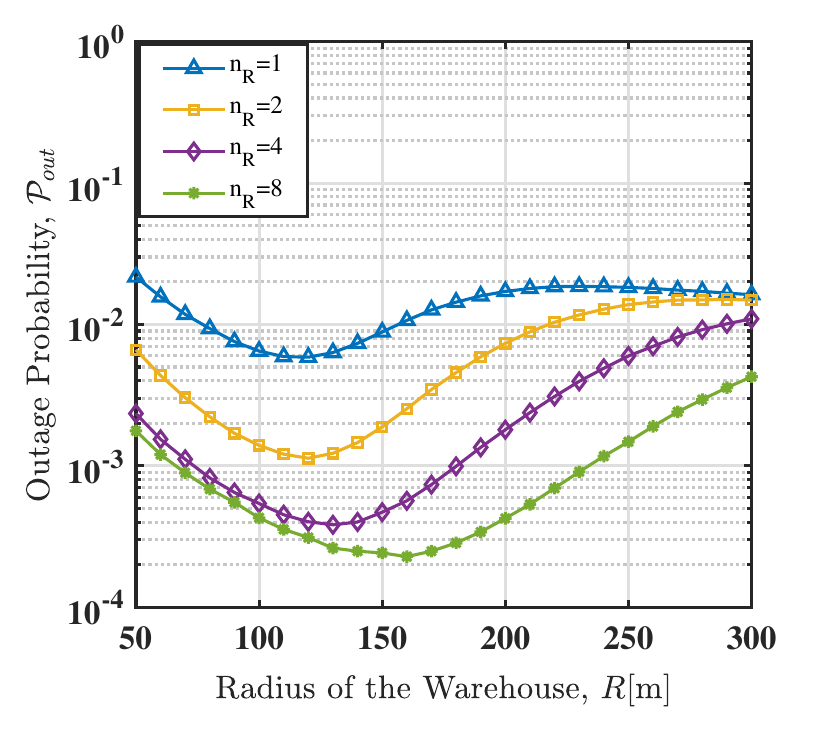}
    \label{fig: R(4)}}
    \hfil
    \subfloat[]
    {\includegraphics[width=0.32\linewidth]{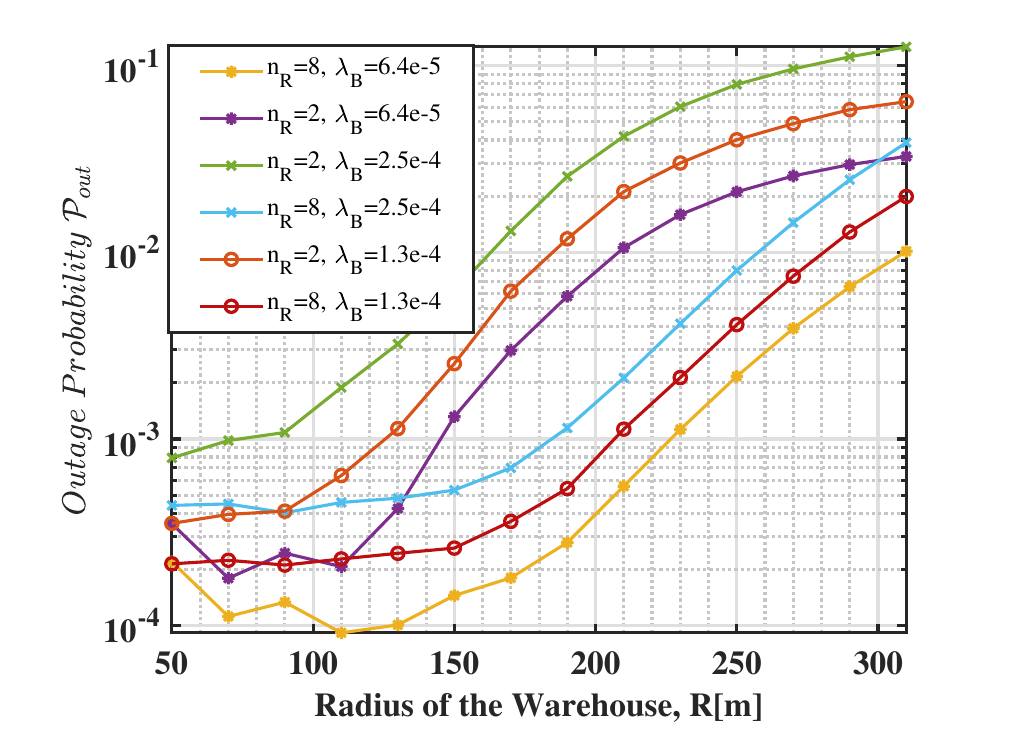}
     \label{fig: R(8)}}
     \caption{Outage probability, $P_{out}$, versus warehouse radius for different numbers of \acp{RIS}, (a) a fixed low number of blockages ($n_{\rm B}=2$), (b) a fixed high number of blockages ($n_{\rm B}=8$), and (c) constant blockage density $\lambda_B = 6.5\times 10^{-5},\;1.3\times 10^{-4},\;\text{and}\; 2.5\times 10^{-4}$ with $R_{ref}=100$\;m, respectively.}
     \label{fig: R}
\end{figure*}
Fig.~\ref{fig: R} illustrates the outage probability $\mathcal{P}_{\text{out}}$ as a function of the warehouse radius $R$ for three scenarios: (a) a fixed low number of blockages ($n_{\rm B} = 2$), (b) a fixed high number of blockages ($n_{\rm B} = 8$), and (c) a constant blockage density $\lambda_B$, for different values of $n_{\rm R}$. For smaller values of $R$, the blockages are relatively denser within the room, resulting in a higher outage probability. As $R$ increases, the blockage density decreases, which leads to a reduction in the outage probability until a critical point is reached. Beyond this point, further increases in $R$ cause the outage probability to rise due to the reduction in signal power caused by the larger distance between the \acp{RIS} and the user.
In Fig.~\ref{fig: R(2)}, when $n_{\rm B} = 2$, the outage probability gradually decreases for $n_{\rm R} = 1, 2, 4,\,\&\,8$. The critical point at which the outage probability starts to increase depends on the number of \acp{RIS}. For instance, the critical point occurs at approximately $R = 100$~m for $n_{\rm R} = 1$, around $R = 120$~m for $n_{\rm R} = 2$, $R = 140$~m for $n_{\rm R} = 4$, and $R = 180$~m for $n_{\rm R} = 8$. This indicates that deploying more \acp{RIS} allows the network to sustain low outage probabilities even as the warehouse deployment region increases.  
In Fig.~\ref{fig: R(4)}, for a higher number of blockages ($n_{\rm B} = 8$), a similar trend is observed, but with higher overall outage levels due to the increased probability of link blockage.\\
In Fig.~\ref{fig: R(8)}, we consider the case when the warehouse radius increases but the blockage density $\lambda_B = \frac{n_B}{\pi R_{\text{ref}}^2}$ is kept constant by scaling $n_{\rm B} \propto R^2$. In this case, the outage probability typically increases with the deployment radius $R$. This is because larger areas lead to greater distances between the \ac{BS}, \acp{RIS}, and the user, which increases the chance of blockages. The figure also shows that deploying more \ac{RIS} nodes (e.g., $n_{\rm R}=8$) improves reliability, particularly in moderate-sized environments. However, as $R$ grows, even multiple \acp{RIS} become less effective due to the rising number of blockages. This highlights that outage performance is influenced not only by blockage density but also by the overall coverage area and the separation between devices. Therefore, in large-scale or densely obstructed deployments, \ac{RIS}-assisted communication systems must be carefully scaled in accordance with both the deployment range and the expected blockage intensity to maintain reliable performance.
\subsection{Effect of the Size of the Blockages}
\label{subsec: Radius of the blockages}
\begin{figure}
    \centering
    \includegraphics[width=0.85\linewidth]{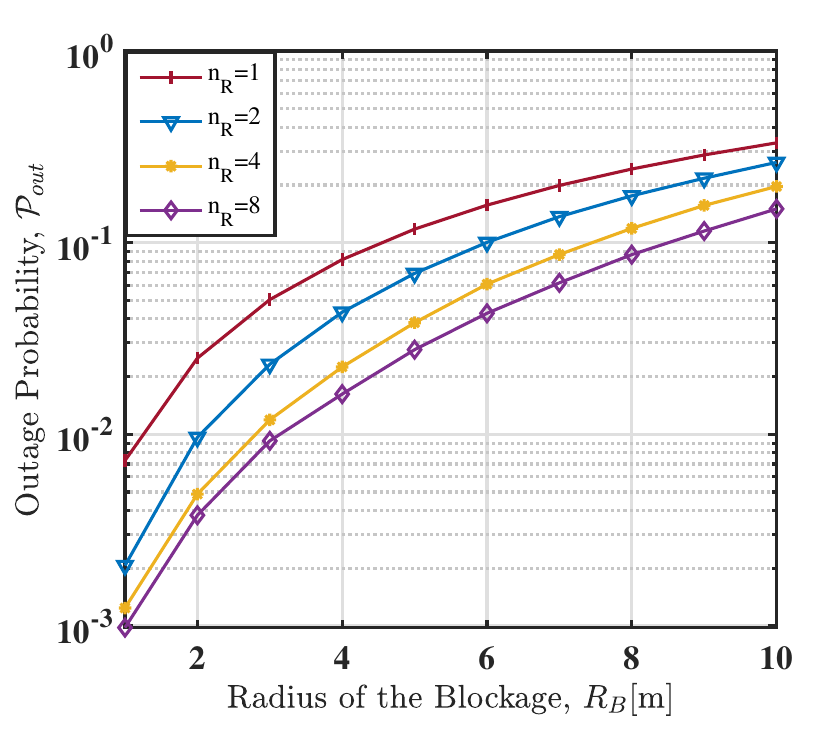}
    \caption{Outage probability, $\mathcal{P}_{out}$, versus the radius of the Blockages$(R_B\;[m])$, for different numbers of \acp{RIS}$(n_R = 1, 2, 4, 8)$, where $n_{\rm B}=4$ and $R=100 \;m$.}
    \label{fig: R_B}
\end{figure}
In Fig.~\ref{fig: R_B}, we plot the outage probability as a function of the radius of the blockages $R_B$ for different values of $n_{\rm R}$. The plot shows an increasing trend as $R_B$ increases. This is because the larger blockages are more likely to block the path between the \ac{RIS}, the \ac{BS}, and the user. When $n_{\rm R}=1$, the outage probability is low for small values of $R_B$ and increases for higher values of $R_B$. The outage probability is low at $R_B=1\;m$ and high at $R_B =10\;m$. When we add more \acp{RIS}, the trend of the plot is the same, but the outage probability is lower compared to a single \ac{RIS} because the \acp{RIS} can reflect the signal around the blockages. For $n_{\rm R} =1$, the outage probability is $10$ times greater than for $n_{\rm R} = 8$ when the size of the blockages is small. However, as the size of the blockage increases, the ratio of the outage probability gradually decreases.
\subsection{Effect of the Number of \acp{RIS}}
\label{subsec: Number of RIS}
\begin{figure}
    \centering
    \includegraphics[width=0.95\linewidth]{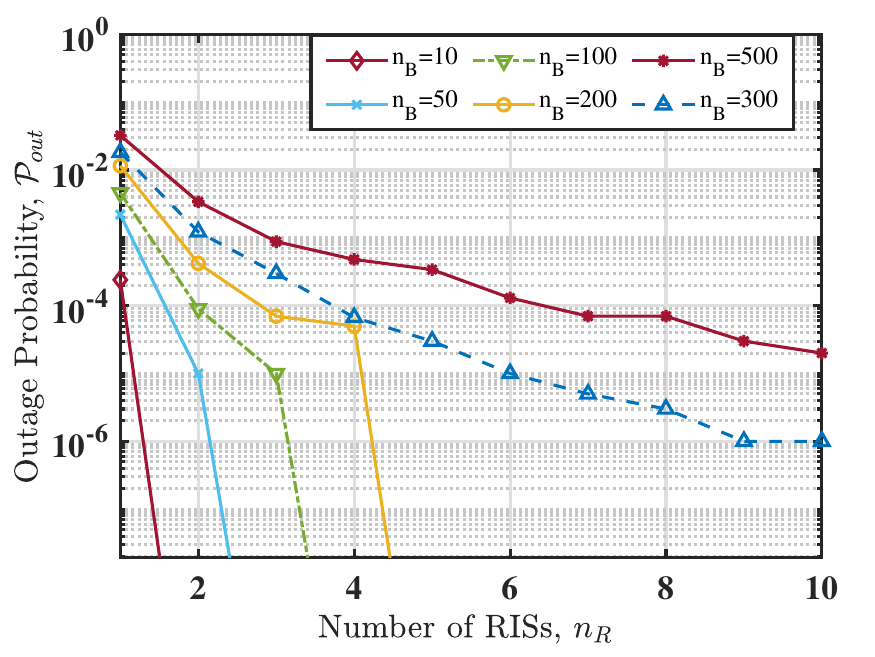}
    \caption{Variation  of the outage probability, $P_{out}$, versus the number of \acp{RIS}, $n_{\rm R}$, for different numbers of blockages $(n_B = 10, 50, 100, 200, 300, 500)$, $R=100 m$ and $R_B=0.2 m$.}
    \label{fig: n_R}
\end{figure}
The plot in Fig.~\ref{fig: n_R} illustrates the impact of the number of \acp{RIS} ($n_{\rm R}$) on the outage probability ($\mathcal{P}_{\text{out}}$) for different values of $n_{\rm B}$. Across all values of $n_{\rm B}$, increasing the number of \acp{RIS} leads to a significant reduction in $P_{\text{out}}$, confirming the effectiveness of \ac{RIS} deployment in enhancing link reliability. For smaller values of $n_{\rm B}$ (e.g., $n_{\rm B} = 10$ or $50$), the outage probability drops sharply with just a few \acp{RIS}, quickly approaching negligible values. This indicates that even minimal \ac{RIS} deployment can yield substantial performance improvements in sparse environments. For intermediate values (e.g., $n_{\rm B} = 100$ and $n_{\rm B} = 200$), the decrease is still significant, but slightly more \acp{RIS} are required to reach very low outage levels. In contrast, for larger values of $n_{\rm B}$ (e.g., $n_{\rm B} = 300$ or $500$), the decrease in outage probability becomes more gradual, with $P_{\text{out}}$ remaining above $10^{-5}$ even with 10 \acp{RIS} in the $n_{\rm B} = 500$ case. This highlights that higher blockage density or complexity requires more \acp{RIS} to achieve comparable reliability. Furthermore, most curves exhibit a saturation trend, where adding more \acp{RIS} beyond a certain point yields diminishing returns. These results imply that the effectiveness of \ac{RIS} deployment is strongly influenced by the system configuration, and careful optimization of \ac{RIS} quantity is necessary, particularly in scenarios with high $n_{\rm B}$, to meet reliability requirements efficiently. Therefore, strategic planning of \ac{RIS} deployment is essential to balance performance with deployment costs, especially in complex environments.
\subsection{Number of \acp{RIS} Required Under Outage Probability Constraint}
\begin{figure}
    \centering
    \includegraphics[width=0.95\linewidth]{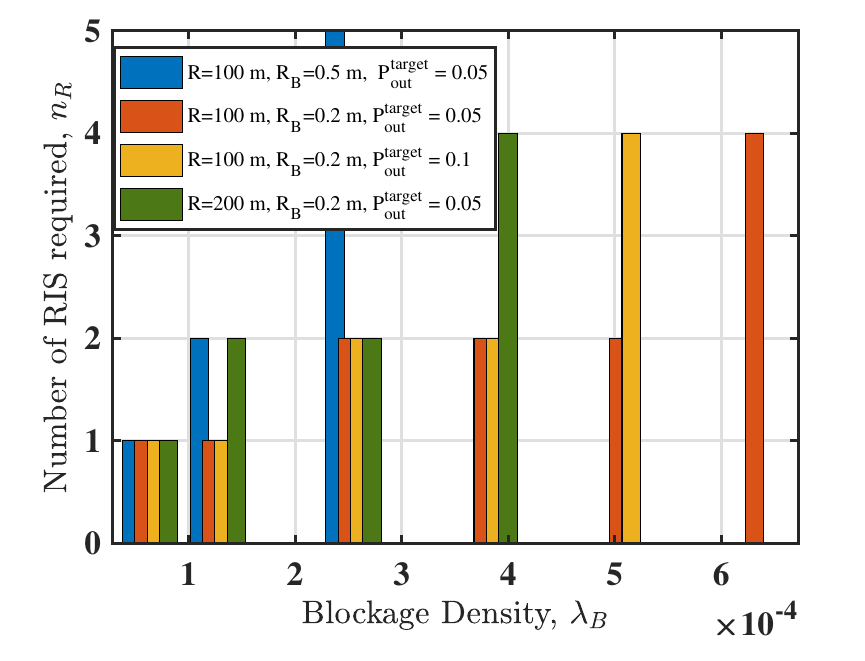}
    \caption{Minimum number of \acp{RIS}, $n_{\rm R}$, required versus blockage density, $\lambda_{\rm B}$, for outage probability targets $P_{\text{out}}^{\text{target}} = 0.05$ and $0.1$, under varying warehouse radii ($R = 100~\text{m}, 200~\text{m}$) and blockage sizes ($R_B = 0.2~\text{m}, 0.5~\text{m}$).}
    \label{fig: lambda_B}
\end{figure}
Fig.~\ref{fig: lambda_B} shows the minimum number of \acp{RIS} ($n_R$) needed to meet a target outage probability ($P_{\text{out}}^{\text{target}}$) for different blockage densities ($\lambda_B$). As $\lambda_B$ increases, more \acp{RIS} are required to overcome link blockages and maintain outage performance.
For example, when comparing scenarios with the same deployment radius $R$ and the target outage probability $P_{\text{out}}^{\text{target}}$ but different blockage sizes (e.g., $R_B = 0.2$~m vs.\ $R_B = 0.5$~m), the case with the larger blockage radius consistently requires more \acp{RIS} across all values of $\lambda_B$. This is because larger blockages occupy more area and are more likely to obstruct communication links.
The influence of the outage probability target is also evident. A more relaxed requirement (e.g., $P_{\text{out}}^{\text{target}} = 0.1$ instead of $0.05$) permits fewer \acp{RIS} to be deployed, as the system can tolerate a higher probability of link failure. Moreover, increasing the deployment radius (e.g., $R = 200$~m vs.\ $R = 100$~m) results in a greater RIS requirement. This is due to both increased link distances and a fixed blockage density $\lambda_B$, the expected number of blockages grows with the area, specifically, by a factor of four when $R$ doubles. As a result, the blockage environment becomes more severe with increasing $R$, necessitating the addition of more \acp{RIS} to maintain reliable coverage.
Therefore, the figure highlights that the blockage size, area dimensions, and the outage probability significantly influence \ac{RIS} deployment, emphasizing the need for careful system design in blockage-prone environments where reliable connectivity is critical.
 \section{Conclusions} 
 \label{sec: Con}
 Placing \acp{RIS} too close or too far from the \ac{BS} increases the likelihood of simultaneous blockages in the \ac{BS}-\ac{RIS} and \ac{RIS}-user links, which degrades link reliability. In this paper, we extended the analysis to networks with multiple \acp{RIS} under realistic blockage conditions by incorporating intra- and inter-link correlation models. We derived closed-form expressions for \ac{LoS}/\ac{NLoS} probabilities and outage probability, revealing how blockage characteristics and user location influence coverage. For a single \ac{RIS}, outage probability increases sharply with blockage density, and we show that beyond a critical number of blockages, traditional relays can outperform a single \ac{RIS}. However, we demonstrated that deploying multiple \acp{RIS} can effectively restore and enhance network performance, offering a scalable solution to combat severe blockage. Our study also shows that assuming independent blockages may lead to inaccurate predictions when obstructions are large, highlighting the importance of accounting for blockage correlation in the system model. Finally, we analyzed the effect of warehouse size and blockage density on rate coverage, identifying practical design thresholds where \acp{RIS} can outperform relaying-based solutions. These findings provide useful insights for practical deployment of reliable, low-latency \ac{RIS}-aided \ac{IIoT} networks capable of meeting stringent \ac{URLLC} requirements. 
\bibliographystyle{ieeetr}
\bibliography{refs_rp}
\end{document}